\documentclass[pdftex]{copernicus}

\begin{document}

\title{Observational features of equatorial coronal hole jets}

\author[1]{G.~Nistic\`o}
\author[2]{V.~Bothmer}
\author[3]{S.~Patsourakos}
\author[1]{G.~Zimbardo}

\affil[1]{University of Calabria}
\affil[2]{Goettingen University}
\affil[3]{Naval Research Laboratory - Also at George Mason University}


\runningtitle{EUV equatorial coronal hole jets}

\runningauthor{Nistic\`o at al.}

\correspondence{G. Nistic\`o\\ (giuseppe.nistico@fis.unical.it)}

\received{}
\pubdiscuss{} 
\revised{}
\accepted{}
\published{}


\firstpage{1}

\maketitle

\abstract{
Collimated ejections of plasma called `` coronal hole jets'' are commonly observed in polar coronal holes. However, such coronal jets are not only a specific features of polar coronal holes but they can also be found in coronal holes appearing at lower heliographic latitudes.
In this paper we present some observations of ``equatorial coronal hole jets'' made up with data provided by the STEREO/SECCHI instruments during a period comprising March 2007 and December 2007. The jet events are selected by requiring at least some visibility in both COR1 and EUVI instruments.
We report 15 jet events, and we discuss their main features. For one event, the uplift velocity has been determined as about 200 km s$^{-1}$, while the deceleration rate appears to be about 0.11 km s$^{-2}$, less than solar gravity. The average  jet visibility time is about 30 minutes, consistent with jet observed in polar regions. On the basis of the present dataset,  we provisionally conclude that there are not substantial physical differences between polar and equatorial coronal hole jets.
}

\introduction
The STEREO (Solar TErrestrial RElations Observatory) mission represents a milestone in the exploration and observation of the Sun and of the inner Heliosphere. It consists of two identical spacecraft launched in October 2006, orbiting at distances of around 1 AU, one ahead of the Earth (STEREO A) and the other behind the Earth (STEREO B) \citep{Russel08}. Their angular separation increases at a rate of about 44 degrees per year with respect to the Sun-Earth line and they provide in situ measurements of the interplanetary medium and images of the Sun in the visible and at Extreme UltraViolet (EUV) wavelengths \citep{Kaiser08}. Solar images are collected by the SECCHI (Sun-Earth Connection Coronal and Heliospheric Investigation) instrument suites \citep{Howard08} on board STEREO A and B. SECCHI contains five telescopes on each satellite. The Extreme-UltraViolet Imager (EUVI) takes full-disk images at 171, 195, 284 and 304 \AA, and the white light (WL) COR (CORonograph)1 and COR 2  take WL pictures of the near-Sun corona. These telescopes comprise the SCIP package pointing towards the Sun, and are  complemented by the off-pointing HI (Heliospheric Imager) 1 and HI 2 observing the heliosphere \citep{Howard08}. 

Solar observations in the EUV allow studies of the dynamics of various phenomena like flares, erupting prominences, coronal waves, amongst others. Interesting examples are coronal hole jets, which are narrow and fast ejections of plasma \citep{Shimojo96}. They are best observed inside polar coronal holes when the plasma ejections are seen in emission against the dark background and are not obscured by bright ambient coronal structures. Early observations of coronal jets were provided by Skylab and, in more detail, later by Yohkoh in the soft X-rays \citep{Shimojo96,Shimojo00}. Polar coronal jets were studied by \citet{Wang98} using images of LASCO (Large Angle and Spectrometric COronagraph) and EIT (Extreme ultraviolet Imaging Telescope) instruments on board SOHO (Solar and Heliospheric Observatory). Recently, also Hinode has provided important data on polar jet parameters like duration, size, speed \citep{Savcheva07,Kamio07,Moreno-Insertis08,Filippov09}.
Typically, jets have a duration of tens of minutes to slightly longer than 1h \citep{Savcheva07,Cirtain07}, lengths of 1--20 $\times$10$^4$ km , thickness of 1--4 $\times$10$^4$ km,  and speed ranging from 150 km s$^{-1}$ to over 800 km s$^{-1}$ \citep{Shimojo96,Cirtain07}. 

These phenomena are thought to be triggered by reconnection process between the large-scale open unipolar magnetic field of the coronal hole and small-scale closed bipoles, emerging from the photosphere \citep{Shibata92}. In most cases, the magnetic bipoles are emerging in coronal holes because of the plasma flows associated with solar granulation. The jet morphologies vary widely; one relatively common shape is the so-called {\it Eiffel Tower} (ET) jet \citep{Yamauchi04}, in which the jet is located on the loop top and both loop legs show brightening (so that the overall shape is reminiscent of the Eiffel Tower); in such a case magnetic reconnection appears to happen on top of the loop; this may be interpreted as reconnection of the large scale coronal hole magnetic field with a magnetic bipole axis  approximately perpendicular to the solar surface ({\it e.g.}, \citet{Filippov09}).

Another common shape is the so-called $\lambda$-type jet \citep{Shibata92,Yokoyama96,Filippov09}, in which the jet appears to be associated with one of the loop legs (at least in the early phase; later, a shift of the jet toward the other leg is sometimes observed). This may be interpreted as reconnection of the large scale coronal hole magnetic field with a magnetic bipole on small-scale with its axis directed approximately parallel to the solar surface ({\it e.g.}, \citet{Shibata92, Filippov09}). Because of the temporal evolution of the magnetic structure, these two morphologies are not mutually exclusive for a given event.
    
The two points of view provided by the twin STEREO satellites are very helpful for the identification of coronal hole jets since it is possible to assess what is the three dimensional (3D) structure of the jet, to understand what projection effects are present in single point observations, and to establish the true 3D velocity for the jet \citep{Patsourakos08}.

A first catalog of polar jets observed by STEREO/SECCHI has been given by \citet{Nistico09}. Here we present a different set of events, that is equatorial coronal jets observed in the period from March to December 2007, during the same period as the polar jet investigation of \citet{Nistico09}. An interesting example of an X-ray equatorial jet was also reported from analysis of Hinode observations \citep{Moreno-Insertis08}. Such jets are found in  equatorial coronal holes, and their observation is often hindered by the presence of other ambient coronal structures, especially coronal streamers. For this reason the number of reported equatorial jets can be substantially smaller than the overall number of observed jets.
For polar jets, \citet{Nistico09} report 79 events, observed from March 2007 to April 2008; here we report 15 equatorial jets observed in the same period. Indeed, we were not able to identify further equatorial jets from December 2007 to April 2008.
   
The criteria according to which coronal jet events were identified and selected from the SECCHI dataset are explained in Section 2, including a brief summary of their basic characteristics, like duration, size, and typical speeds. Section 3 shows some typical examples of equatorial jets, while the conclusions are given in Section 4. 

\begin{figure*}[t]
\begin{center}
\vspace*{1mm}
\begin{tabular}{c c c }
\includegraphics[width=5.8 cm]{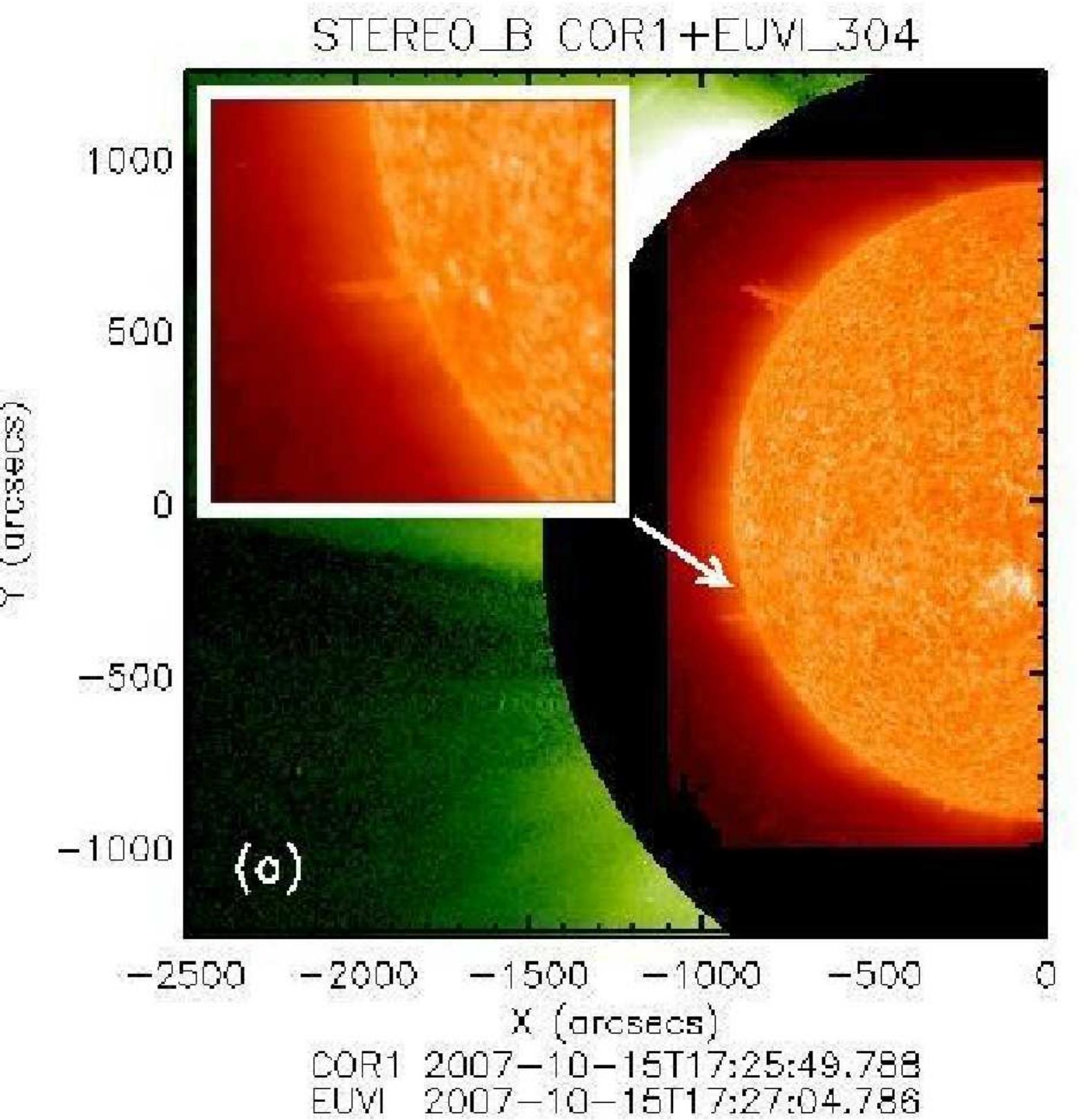} & 
\includegraphics[width=5.8 cm]{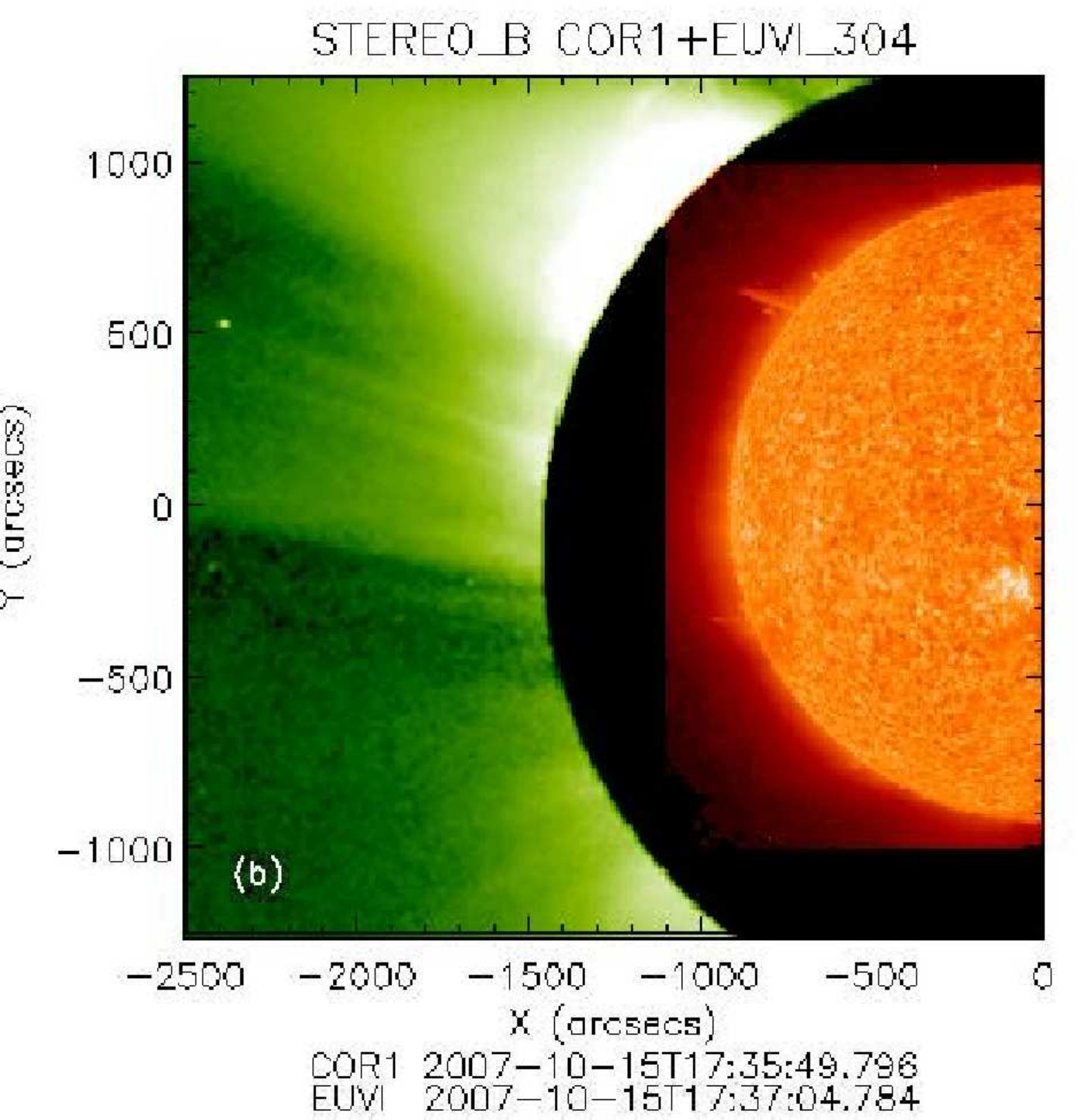} &
\includegraphics[width=5.8 cm]{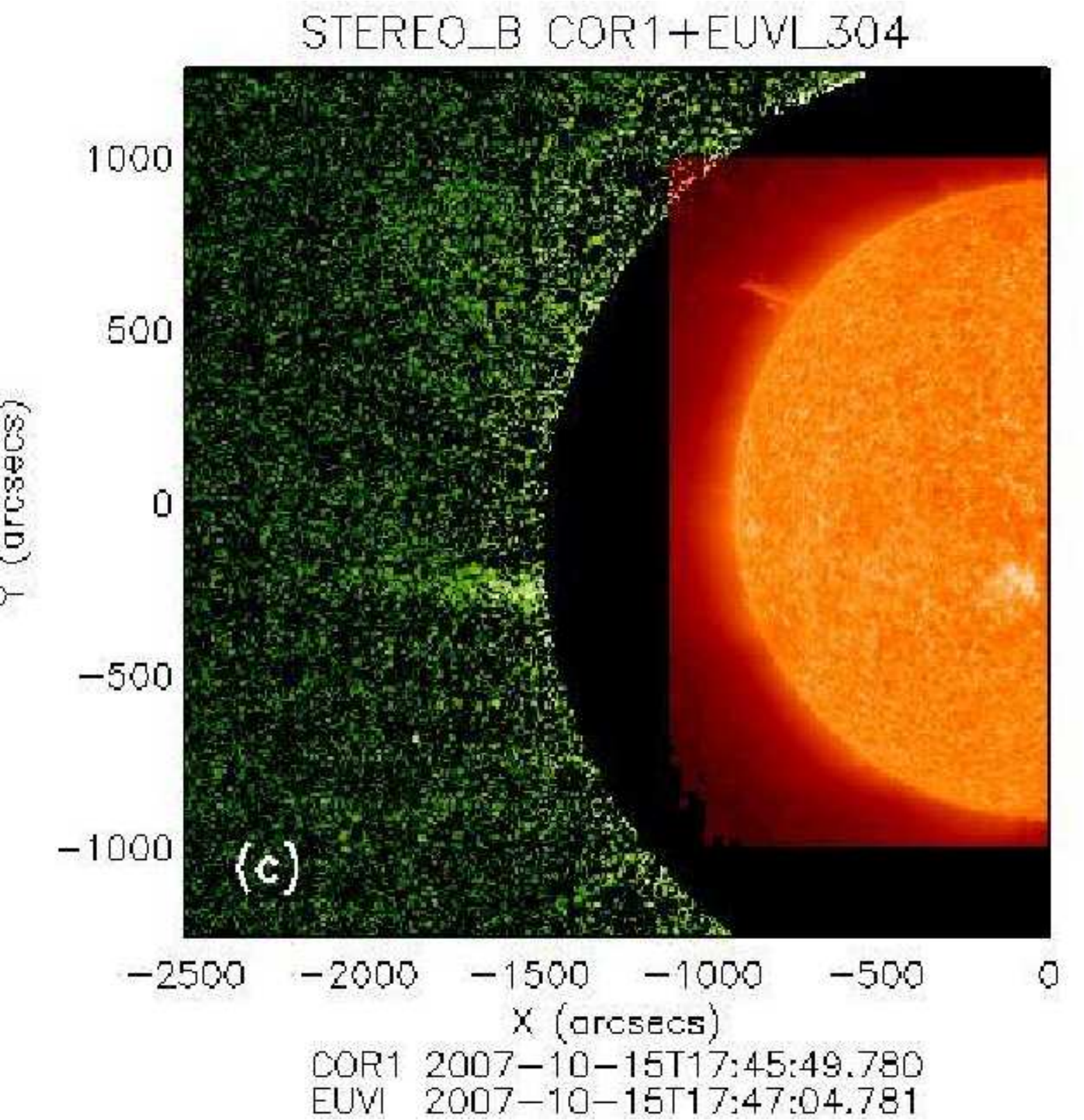} \\
\end{tabular}
\caption{Example of an EUV equatorial jet and its counterpart in the COR 1 field of view as seen by STEREO B on 15 October, 2007.}
\label{fig_1}
\end{center}
\end{figure*}

\section{Selection criteria}

It is usually assumed that a jet is the result of magnetic reconnection happening in the solar corona between the large-scale open unipolar magnetic field of the coronal hole and small-scale closed magnetic field of a bipole \citep{Shibata92, Yokoyama96, Moreno-Insertis08, Pariat09}.
Following \citet{Nistico09}, the identification of equatorial jets has been performed using on-line data available at \url{http://secchi.nrl.navy.mil/} and inspecting daily EUVI movies (\url{http://cdaw.gsfc.nasa.gov/stereo/daily movies/}) and the COR 1 movies (\url{http://cor1.gsfc.nasa.gov/dailymov/MPEG/}). For the selection of events we required the events to show up both in the EUVI and COR 1 fields of view, with the purpose of distinguishing jets from other phenomena like spicules or macruspicules which do not show an ejection of matter in the outer corona. In practice, jets where visually identified first in COR 1 and then in EUVI. Since many events are rather faint, for all reported events we checked for evidence in the COR 1 difference images.    
Differently from \citet{Nistico09}, we relax the requirement that each event should be visible by both STEREO A and STEREO B, due to the intrinsic difficulty in observing equatorial jets. Indeed, an event which is near to the limb for one spacecraft, may be behind the limb for the other one, or just on the Sun disk and difficult to observe. Moreover, the detection of equatorial jets is complicated for other reasons, {\it e.g.}: helmet streamers could hinder the visibility of jets in corona; only few short-lived equatorial coronal holes are present during the solar minimum phase. For these reasons, it is obvious that the number of reported equatorial jets is smaller than the number of polar jets occuring in the same time period of observation. 

In Fig. \ref{fig_1} we show some  images of an equatorial coronal hole jet that occurred on 15 October, 2007. The event was observed only by STEREO B. Two helmet streamers are visible in the COR 1 Field-Of-View (FOV), approximately in the north-eastern and south-eastern direction, and a small dark region is distiguished between them. It corresponds to an equatorial coronal hole near the East limb of EUVI, which indicates open magnetic field lines, in contrast to closed ones of helmet streamers (Fig. \ref{fig_1}b). An ejection is noted coming from the coronal hole at 17:16, perfectly visible at 17:27 in the EUVI FOV (Fig. \ref{fig_1}a) and, finally, its counterpart becomes visible in COR 1 at 17:46 (Fig. \ref{fig_1}c). From the times of appearance in EUVI and COR 1, we can estimate a propagation time of about 30 minutes. Considering that the field of view of COR 1 corresponds to 1.4 R$_\odot$ from the Sun center, we can make a lower estimate of the jet speed of 160 km s$^{-1}$. If the same estimate is done with the observation times reported in Table \ref{table_1}, that is considering the delay between the jet appearance in EUVI and COR 1, typical speeds in the range  of 100--200 km s$^{-1}$ are obtained, comparable to those obtained for polar coronal holes \citep{Cirtain07, Savcheva07, Nistico09}. This jet example provides an idea of the methodology used for the detection of equatorial jets.

\section{List of equatorial coronal jets}
\begin{figure*}[t]
\begin{center}
\vspace*{1mm}
\begin{tabular}{c c c }
\includegraphics[width=5.8 cm]{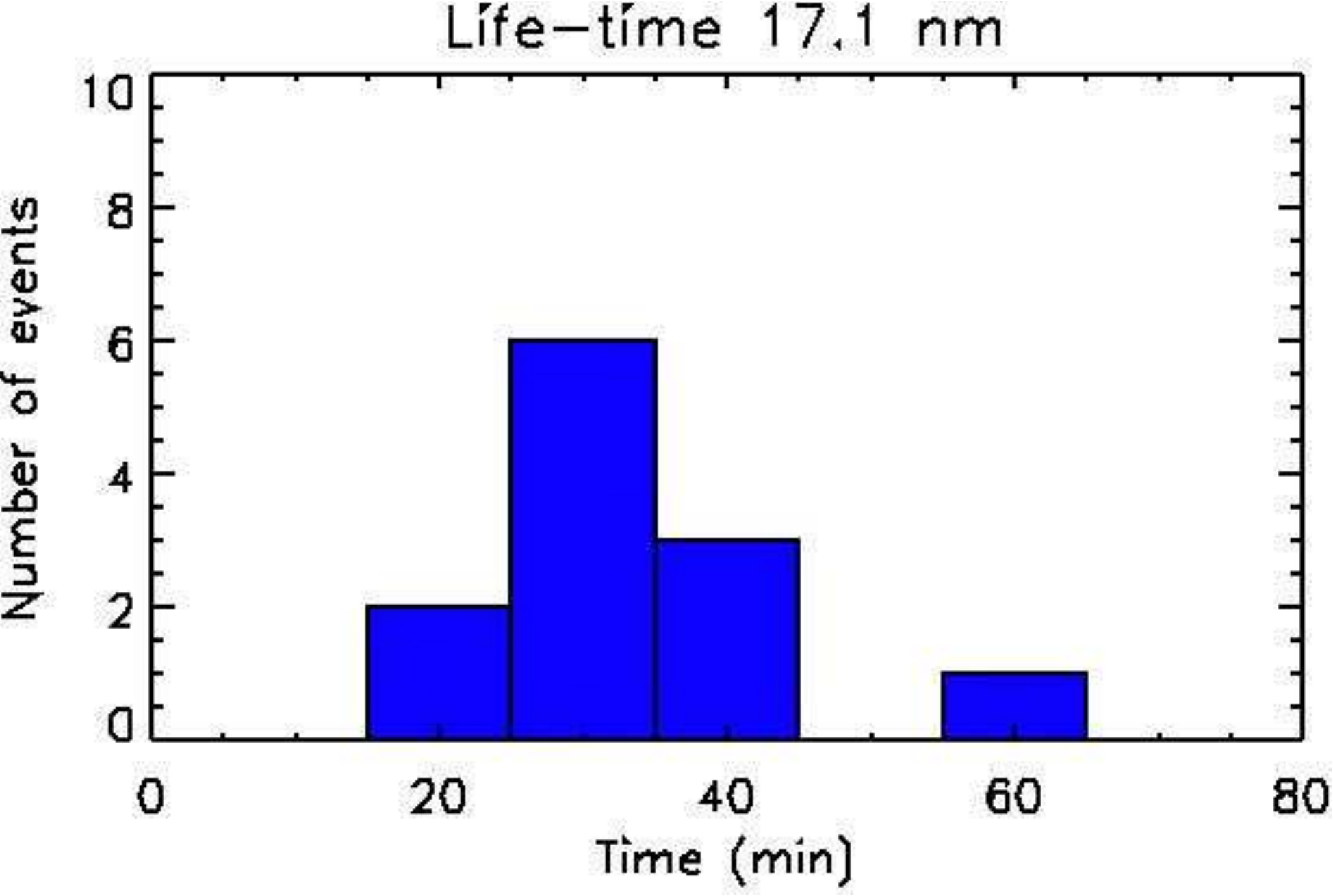} & 
\includegraphics[width=5.8 cm]{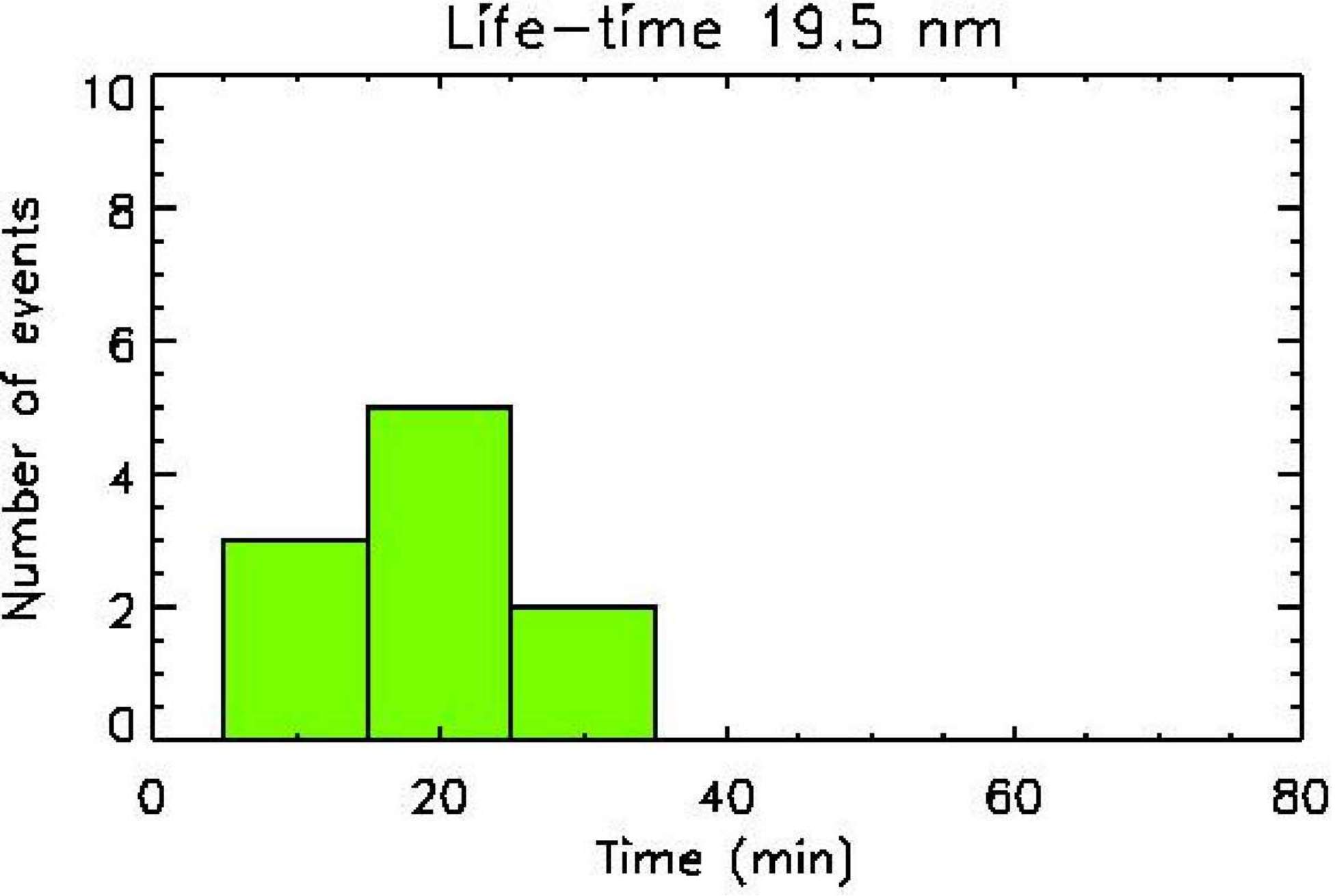} &
\includegraphics[width=5.8 cm]{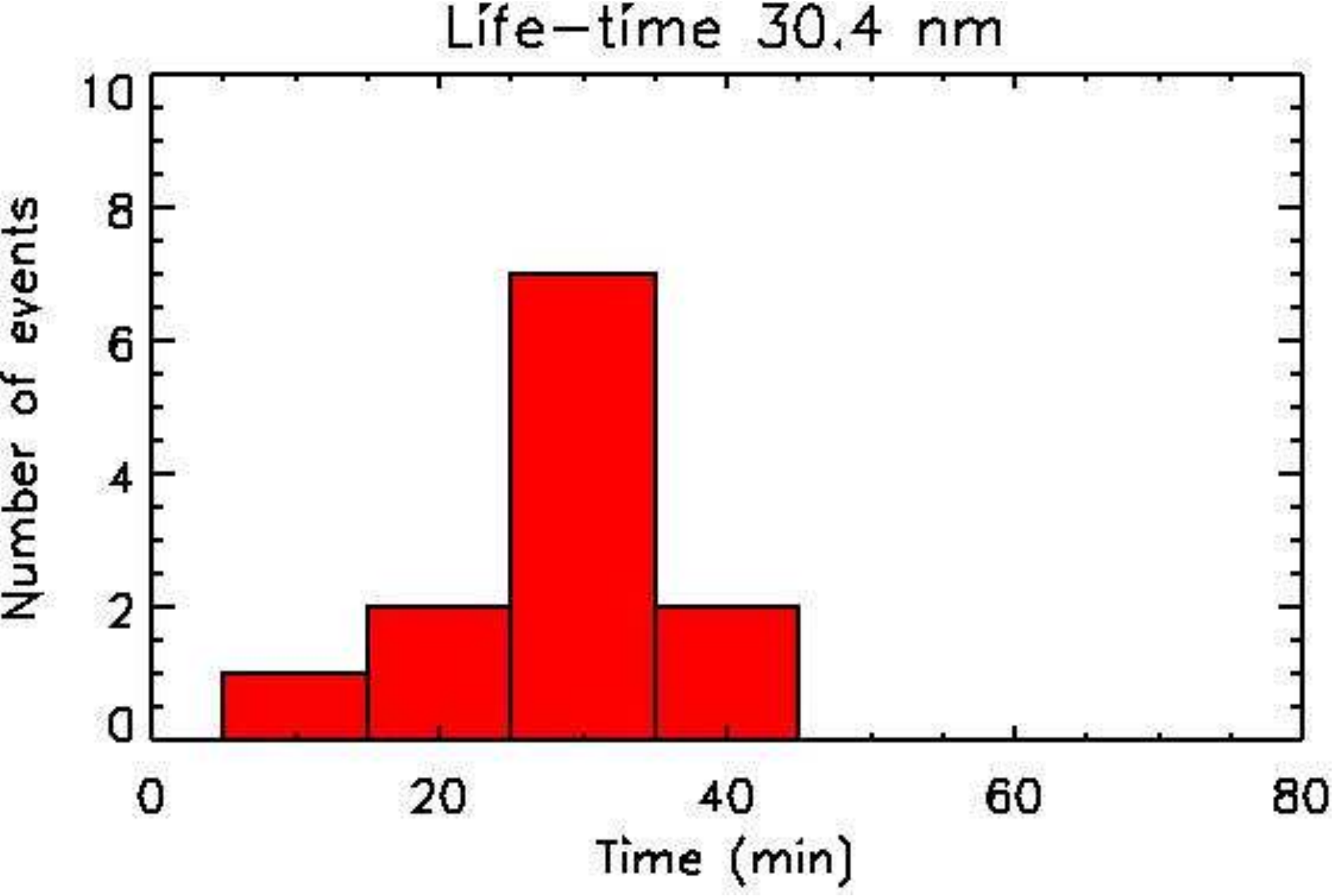} \\
\end{tabular}
\caption{Frequency distribution of the lifetimes of equatorial coronal hole jets at 171, 195, and 304 \AA~averaged in bins of 10 minutes.}
\label{fig_2}
\end{center}
\end{figure*}

Here we present a list of equatorial coronal hole jets identified in the STEREO data during the interval March to December 2007 and discuss their basic properties. The list is given in Table \ref{table_1}; in each column we specify the date, the wavelength of observation in the EUVI, the observation time in the EUVI and COR 1 instruments for STEREO A and B, and the position angle at the solar limb in the EUVI ($\alpha$) and COR 1 FOV($\beta$). Of these 15 identified events, only a few ({\it e.g.}, events N$^\circ$ 9, 10, 11, 12, 14) show a  clear eruption and subsequent propagation in the coronagraph FOV, whereas others ({\it e.g.} events N$^\circ$ 1, 2, 4, 7, 8 of the Table \ref{table_1}) showed a signal in the COR 1 normal or difference images, corresponding to the position of a bright-point (BP) or a very small ejection evolving in time that was noted in the coronal hole.
The jet lifetimes in EUV can be obtained from the observation times reported in Table \ref{table_1}. These data have been organized as histograms for each of the different EUV wavelengths, and are shown in Fig. \ref{fig_2}. It can be seen that the lifetime frequency distributions are similar to those reported for polar jets in X-rays \citep{Cirtain07} and in EUV \citep{Nistico09}, even though the smaller number of observed events is small and despite the observing difficulties of these faint events.   

In the following we show some of the best jet examples and discuss their main features.     
\begin{figure}[t]
\vspace*{2mm}
\begin{center}
\includegraphics[width=8.3 cm]{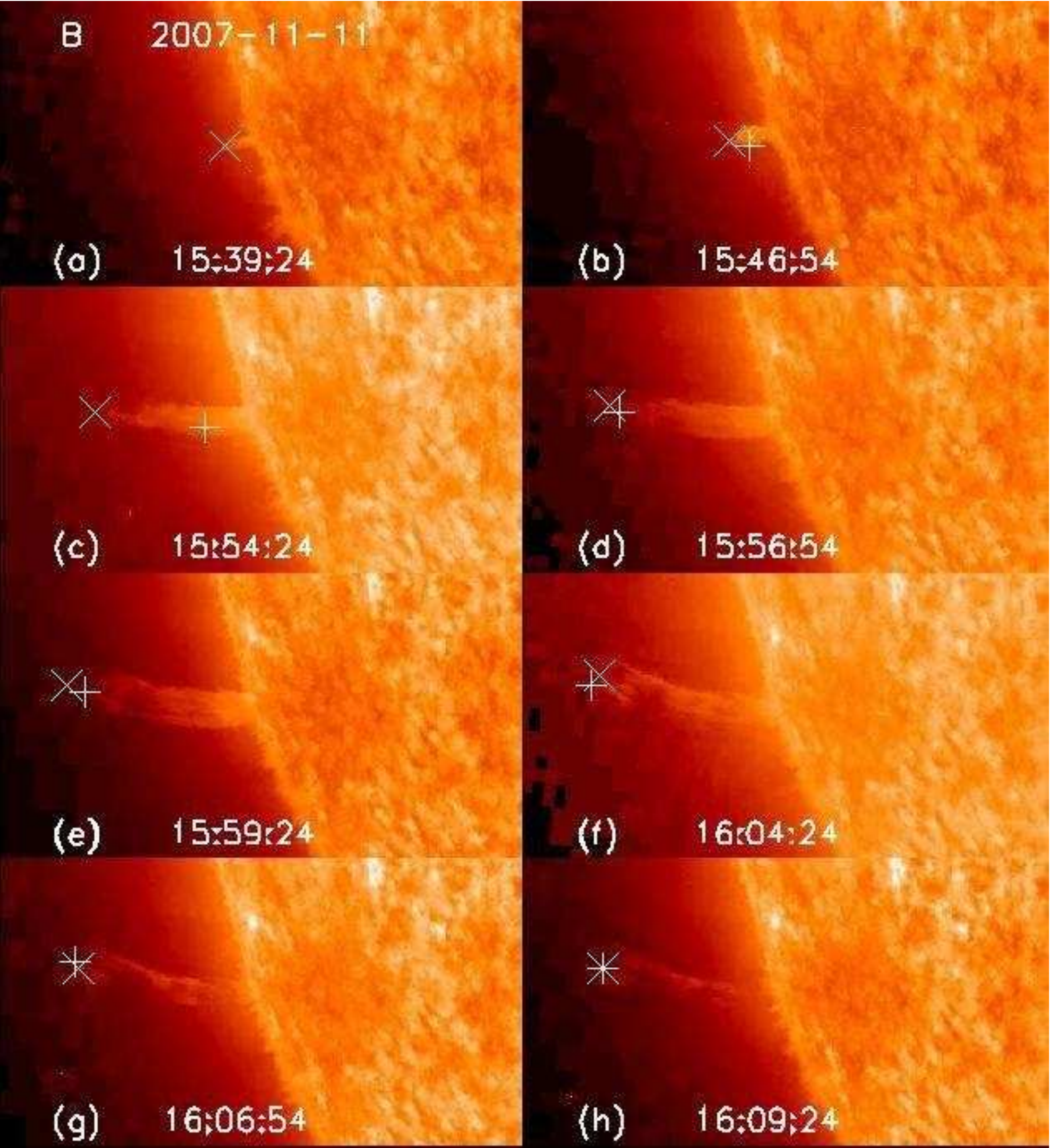}
\end{center}
\caption{Equatorial jet on 11 November 2007 seen by STEREO B at 304 \AA. In each frame, the cross indicates the current position of the leading edge, while the plus sign indicates the same position in the previous frame.}
\label{fig_3}
\end{figure}

\begin{figure}[t]
\vspace*{2mm}
\begin{center}
\includegraphics[width=8.3 cm]{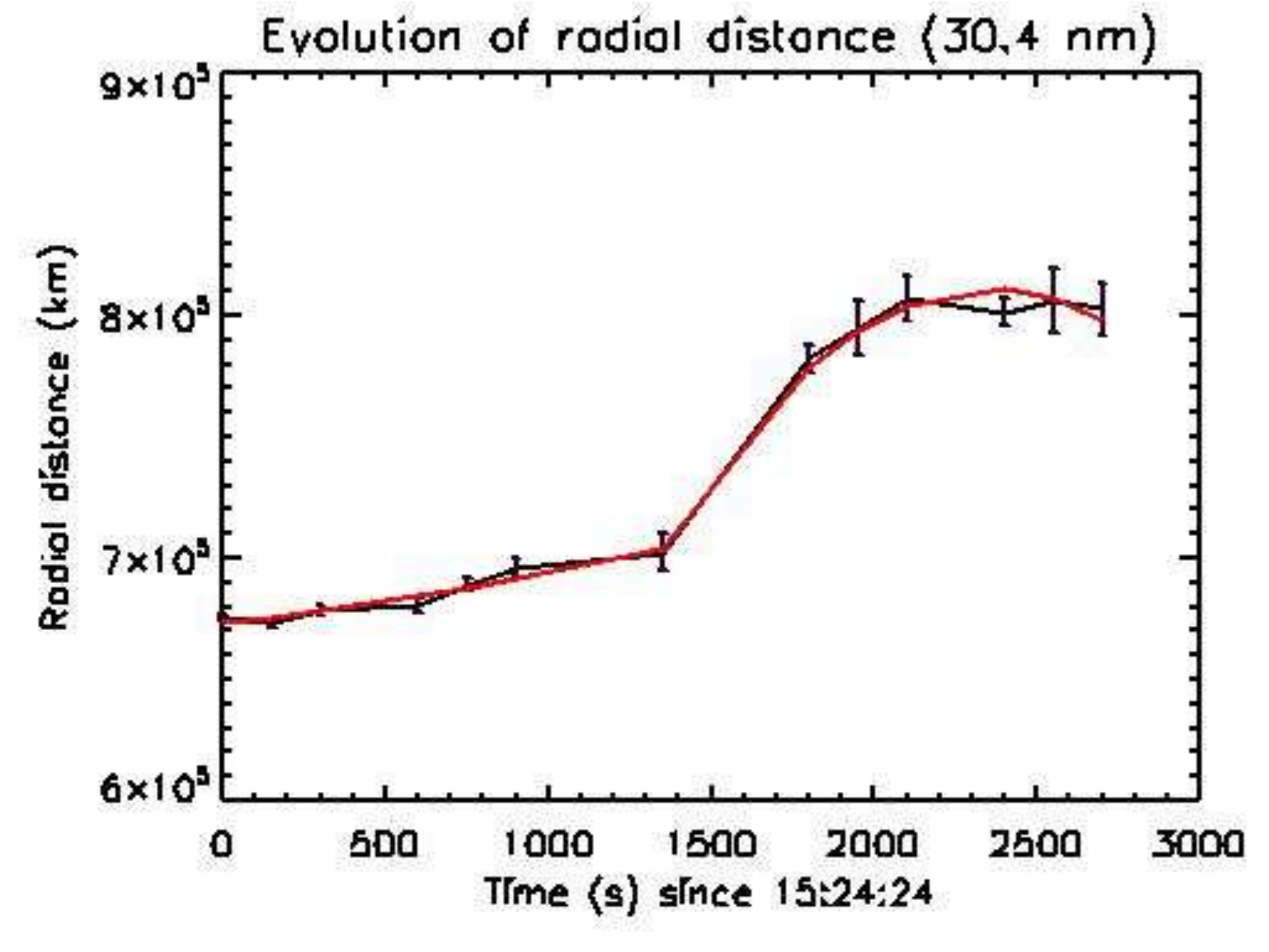}
\end{center}
\caption{Time profile of the jet's leading edge radial distance on 11 November, 2007 seen by STEREO A. The black line represents the positions at each temporal time with its corresponding error bars. The red line is the fit to the data points.} 
\label{fig_4}
\end{figure} 

Fig. \ref{fig_3} shows a very nice example of an equatorial coronal hole jet. It occurred on 11 November, 2007 (event N$^\circ$ 10, this event was also reported  by \citet{Nistico09}). The coronal hole is located on the east limb and is visible only by STEREO B. The coronal hole areas are darker than the quiet Sun areas but overall the corona appears very inhomogeneous, also reflected by the presence of many bright points. Through circles and labeled by BP we have marked the presence of a bright point that increased its brightness in time as seen in STEREO A images, and for which the corresponding jet was visible at the limb with STEREO B.
 Ejection is clearly seen at 304 \AA~and was  propagating through the low corona. First, a very small feature appears on the boundary of the coronal hole, similar to a spicule (Fig. \ref{fig_3}a), which substantially grows in length and size the next 15 minutes (Fig. \ref{fig_3}b-d). After that time a less collimated distribution of emitting material is observed (Fig. \ref{fig_3}f).  From the time of maximum development, corresponding to (Fig \ref{fig_3}e), we can estimate a jet length of 10$^5$ km and a thickness of 2.5$\times$10$^4$ km. These estimates are consistent with the range of sizes found for studied polar jets in the literature (see Introduction). We further studied the eight-time profile of the jet, obtained with the SolarSoft routine (command "\texttt{cursor}") allowing us to determine the position of the leading edge on a map object (this routine also allows estimation of the error in the position by several measurements). We point out that the determination of the jet position might be influenced by plasma cooling during the jet lifetime, which affects the plasma emission. Indeed, the brightness of the jet appears to dim during the time interval 15:59--16:09 UT (Fig. \ref{fig_3}e) from (Fig. \ref{fig_3}h). Determination of the leading edge jet position by neglecting cooling effects seems appropriate in this event because the emitting region maintains its overall shape.  As can be seen in Fig. \ref{fig_4}, the radial position suddenly grows within 7.5 minutes (between 15:45:54--15:54:24 UT). The slope of the curve during this time-interval yields an estimate of the average speed of about 180 km s$^{-1}$. We fitted the points of the slope for two different time intervals with two quadratic poynomials, in order to calculate the initial speed and the phase of constant acceleration. Clearly, in doing such fits we assume a constant force per unit mass, which is unlikely to be precisely true. The first fit, before the impulsive growth of the jet (until 1400 s), yields an initial speed $v_{01}\sim16.93$ km s$^{-1}$ and an acceleration $a_{01}\sim0.00043$ km s$^{-2}$. From the second fit (1400 to 2700 s), we obtain  $v_{02}\sim213.94$ km s$^{-1}$ and $a_{02}\sim-0.11$ km/s$^{-2}$. As can be seen from Fig. \ref{fig_4}, the fitting curves  are within the error bars. A similar two-phase height-time plot was found for the polar jet analyzed by \citet{Patsourakos08}. Such a development can be interpreted as a first slow phase corresponding to a plasma instability build-up, whereas the second phase could correspond to the impulsive relaxation of the build-up energy, triggering  the impulsive eruption of the jet. The acceleration determined for the second phase is negative, indicating a deceleration of  ejected material at higher coronal altitudes  (as can be observed in the images (Fig. \ref{fig_3}f)), however being smaller, in absolute value, than the solar surface gravity ($g_\odot\sim0.270$ km s$^{-2}$). Even taking into account possible projection effects, which cannot be eliminated because this event is only seen by STEREO B, this acceleration discrepancy suggests that some other force, beyond gravity,  (either the pressure gradients or $\bf{j} \times \bf{B}$  force) has accelerated the plasma. The obtained speeds and acceleration are similar to those found for a larger sample of jets discussed by \citet{Nistico09}. In particular the downward acceleration for a polar jet (event N$^\circ$ 15 of that catalogue) was found to be $\sim -0.160$ km s$^{-2}$, leading to similar inferences on the presence of additional forces beyond gravity.  


\begin{figure}[t]
\vspace*{0mm}
\hspace*{0mm}
\begin{tabular}{l l }
\includegraphics[width=4.2 cm]{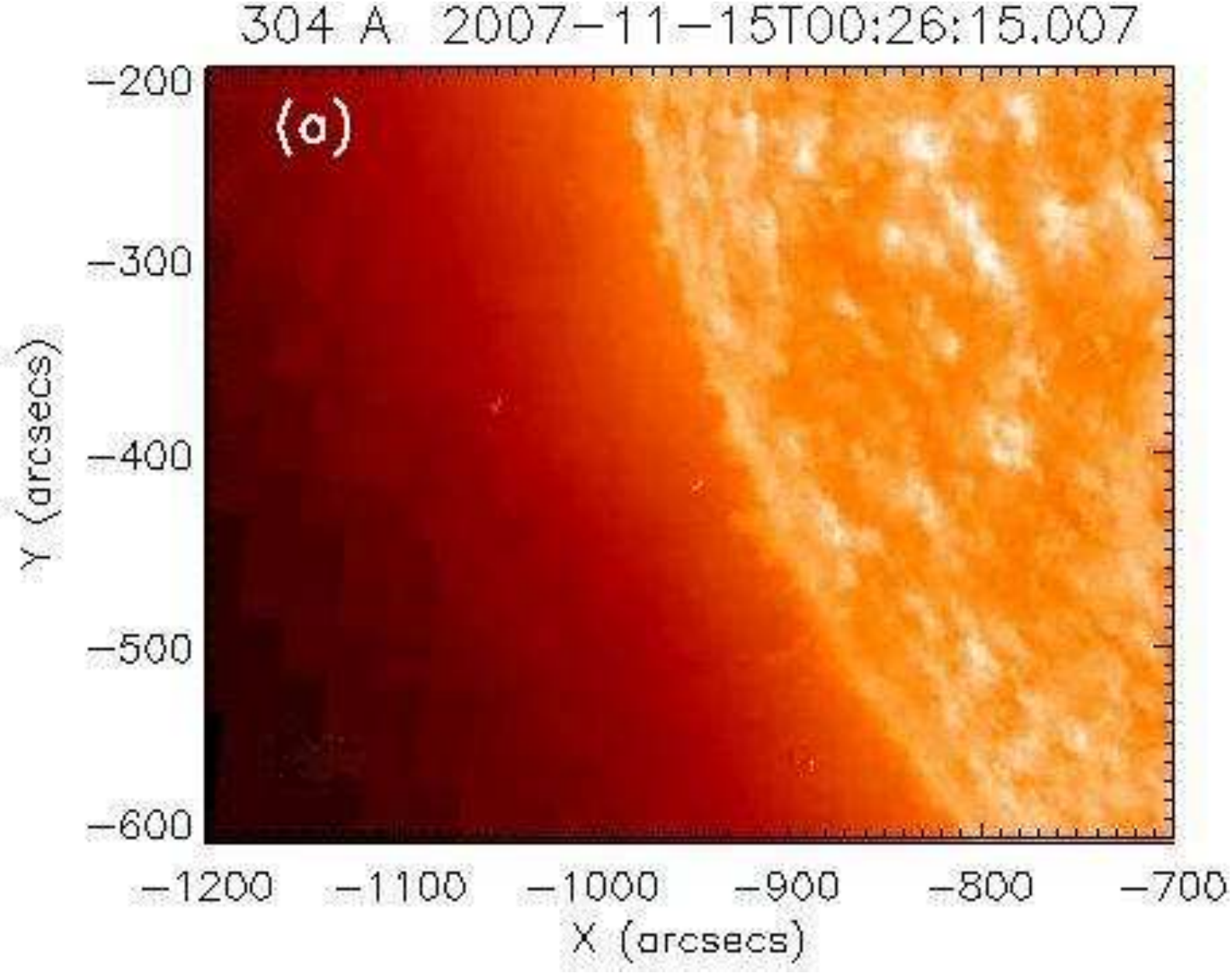} & 
\includegraphics[width=4.2 cm]{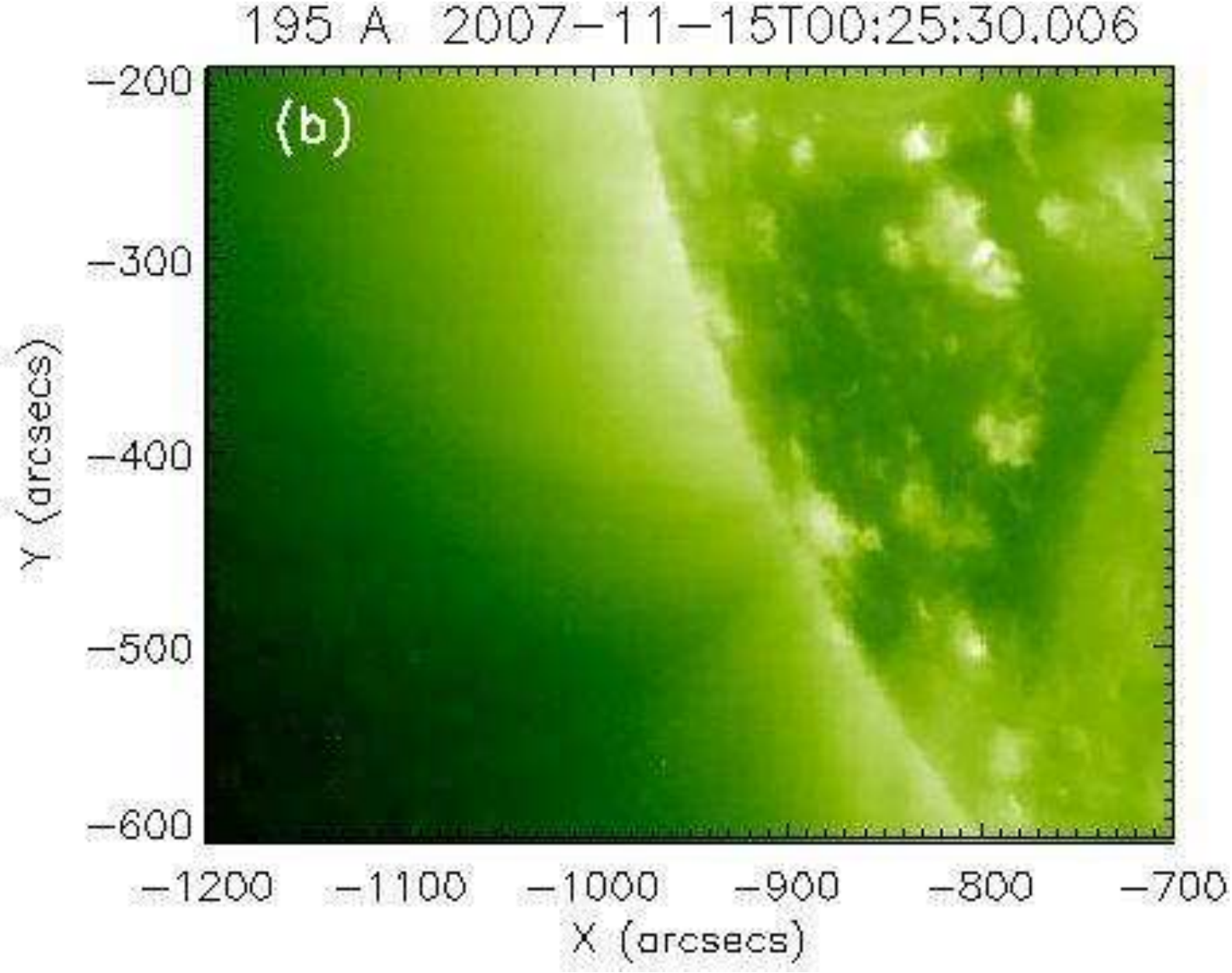}\\
\includegraphics[width=4.2 cm]{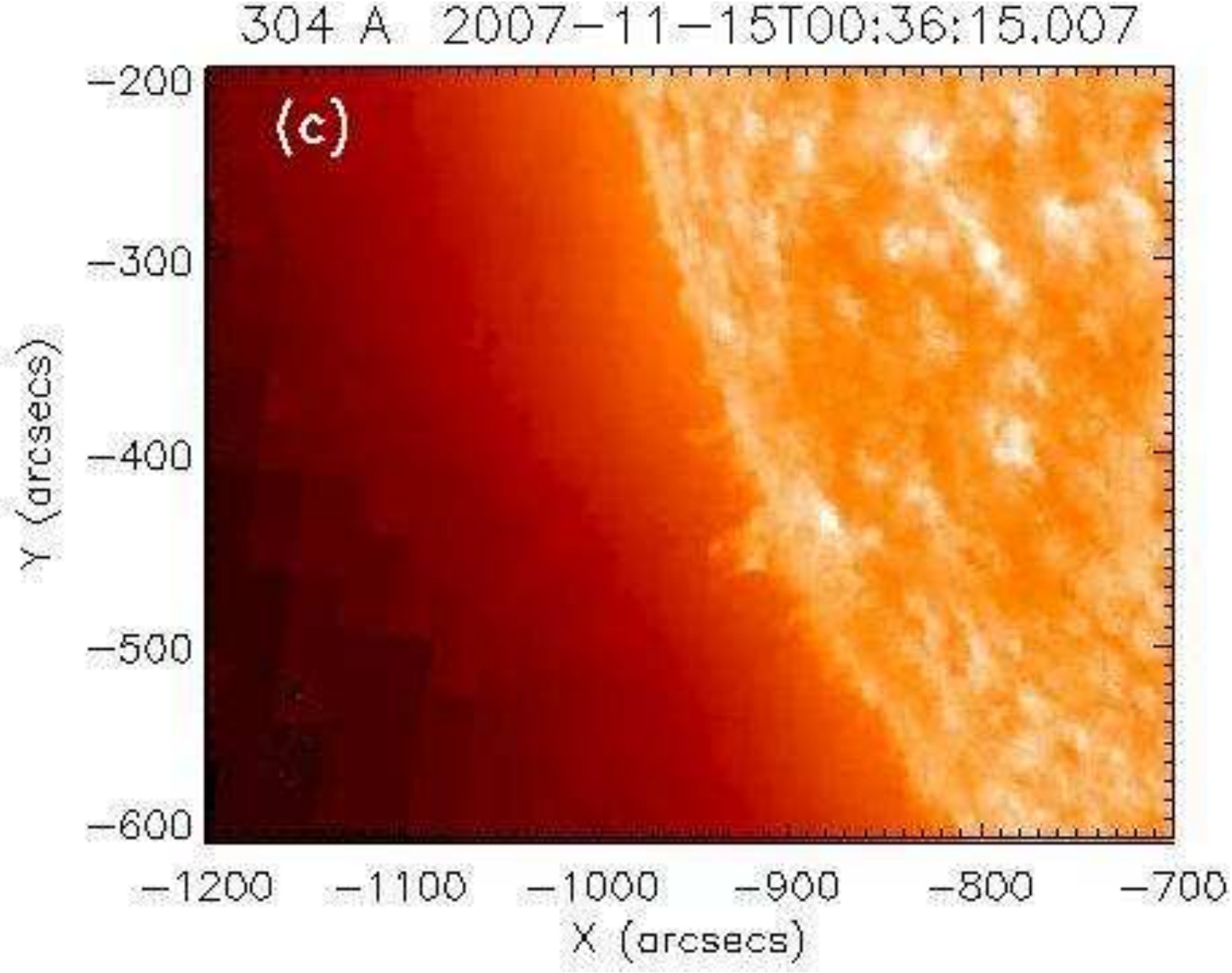} &
\includegraphics[width=4.2 cm]{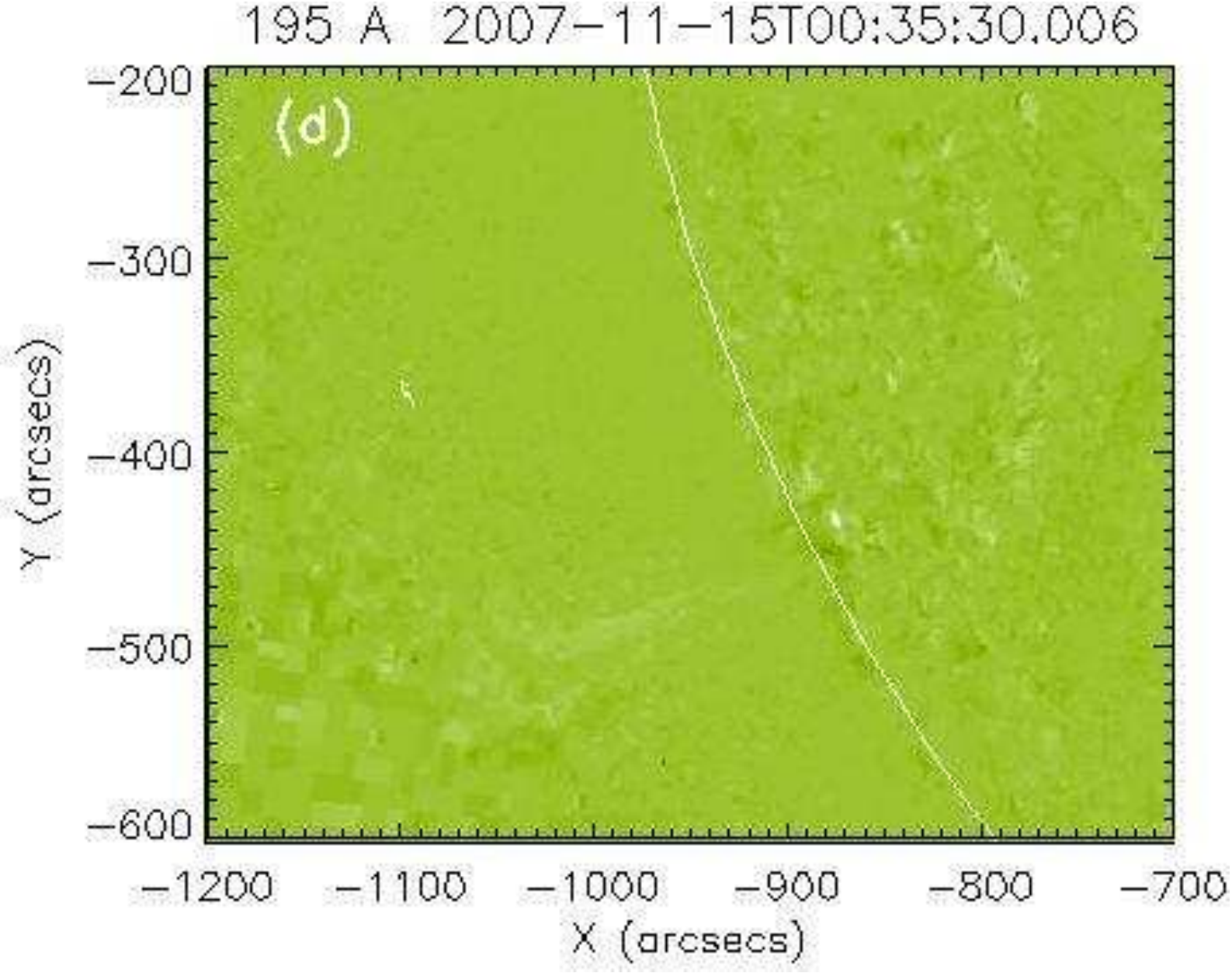} \\ 
\includegraphics[width=4.2 cm]{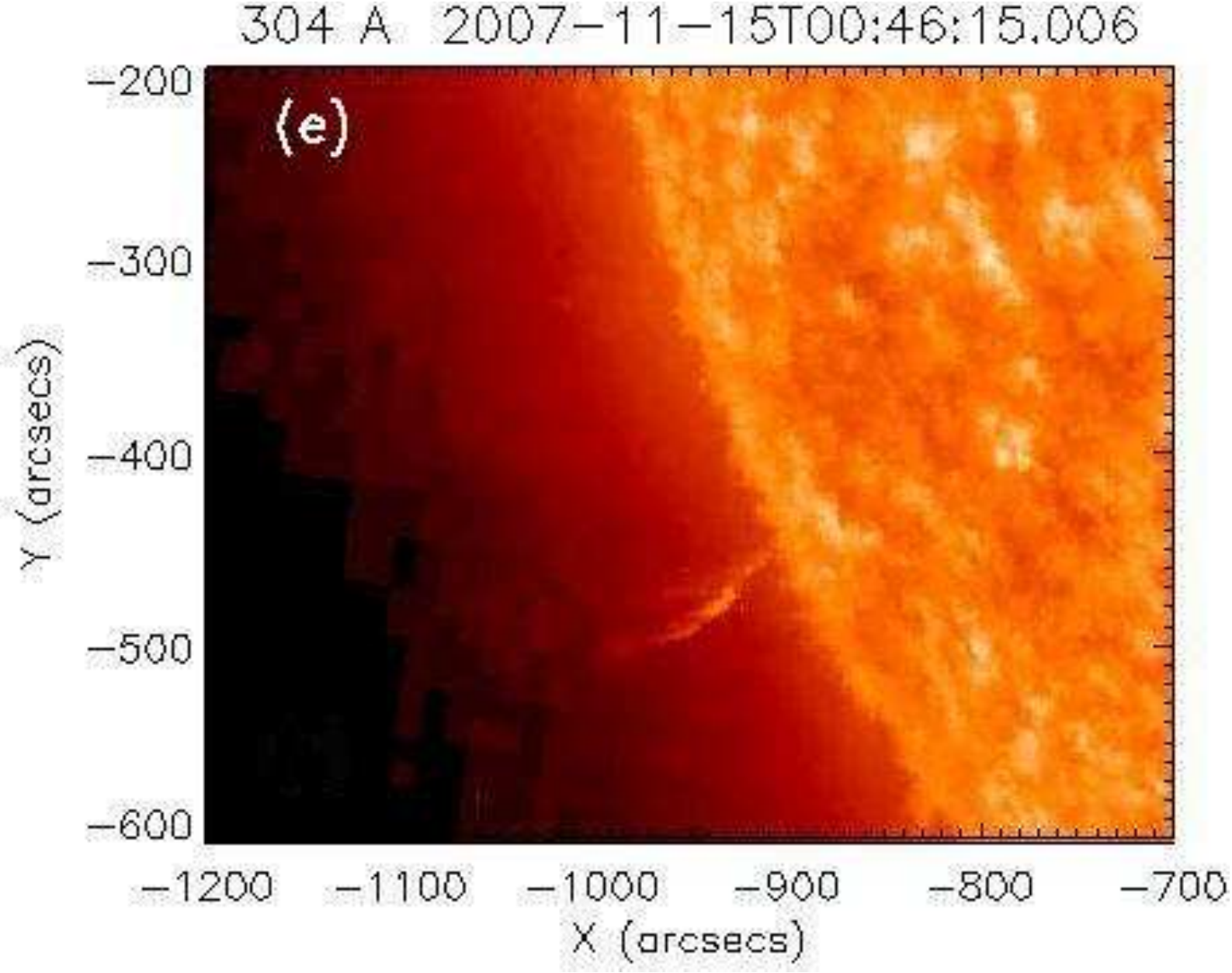} & 
\includegraphics[width=4.2 cm]{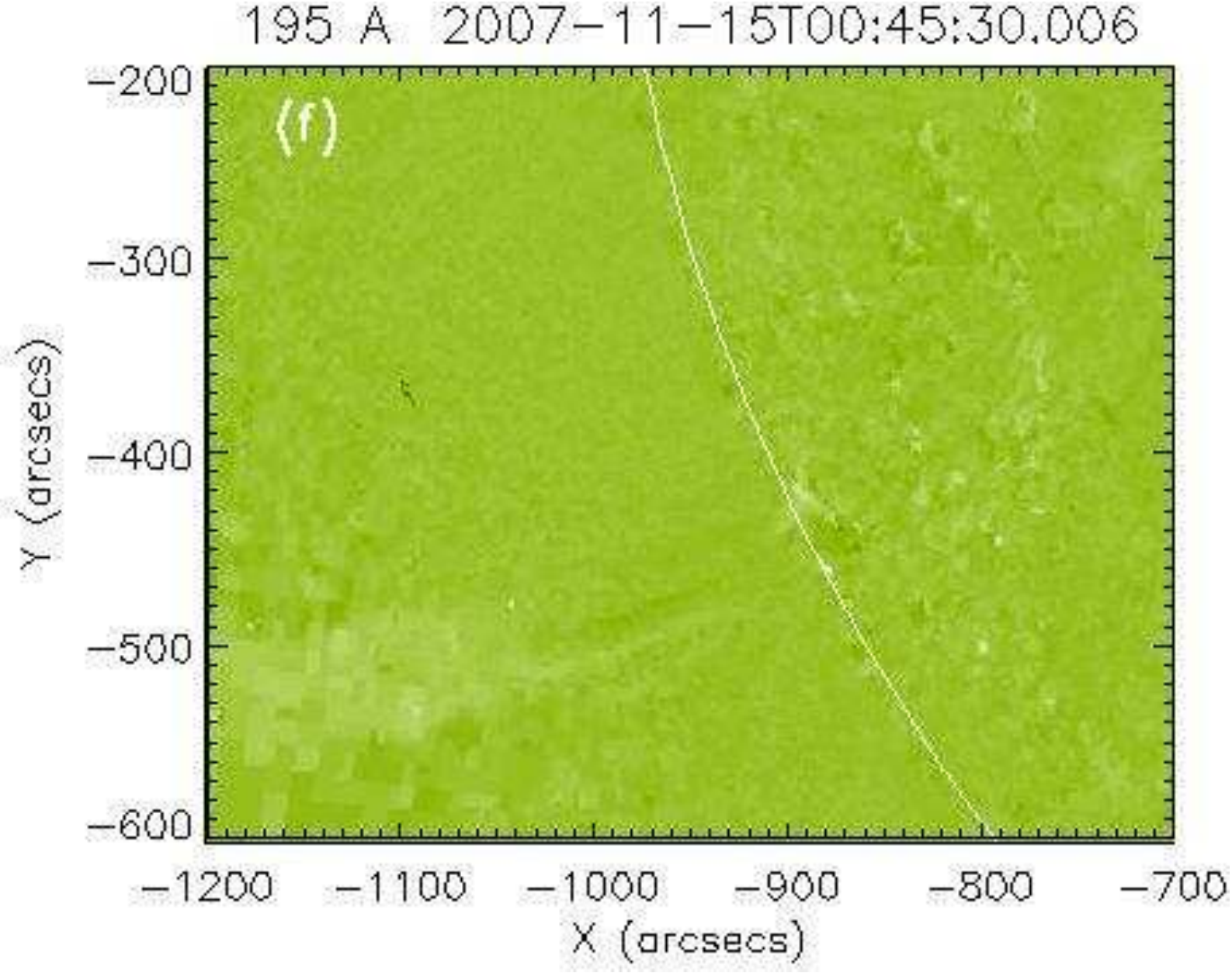} \\ 
\includegraphics[width=4.2 cm]{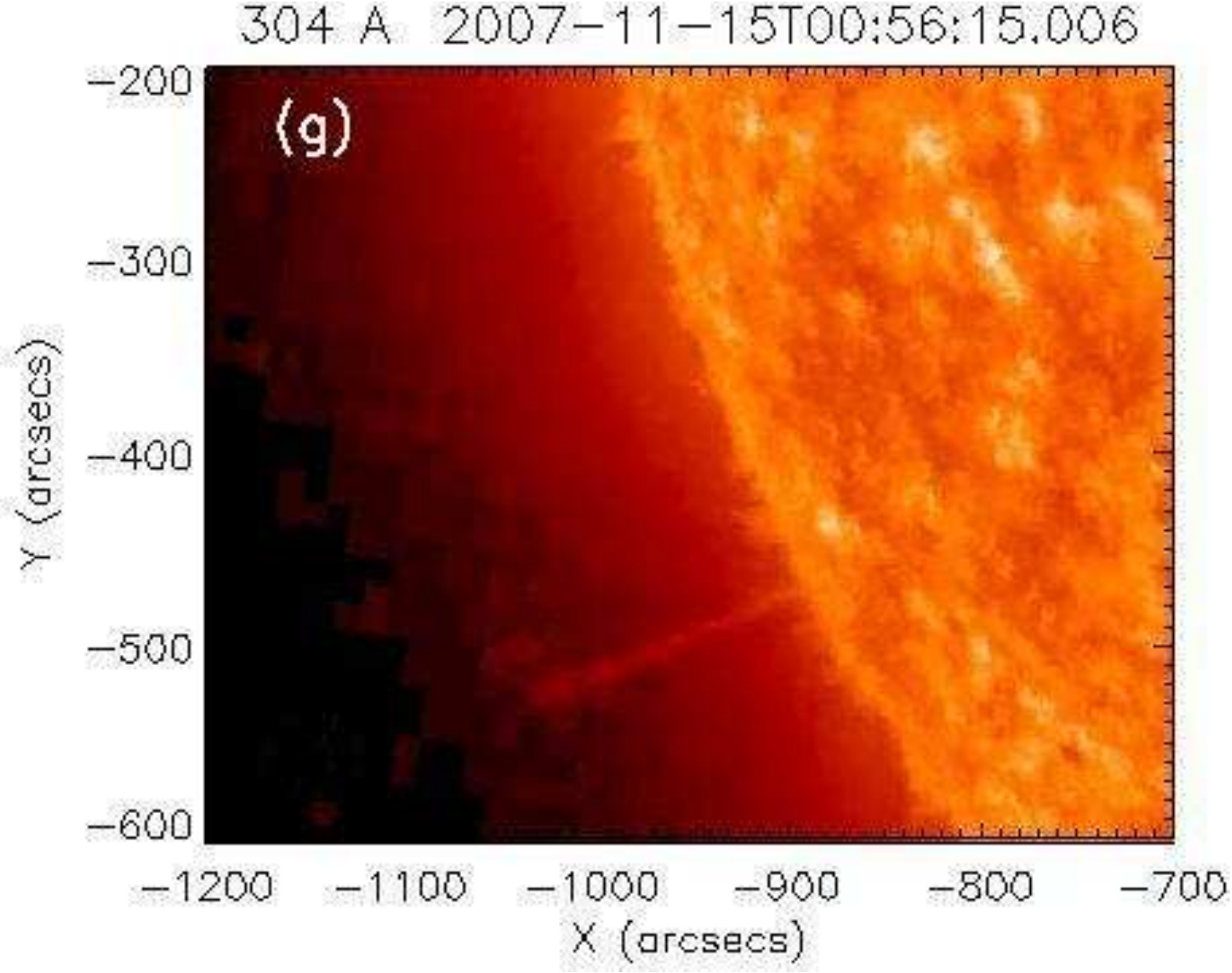} & 
\includegraphics[width=4.2 cm]{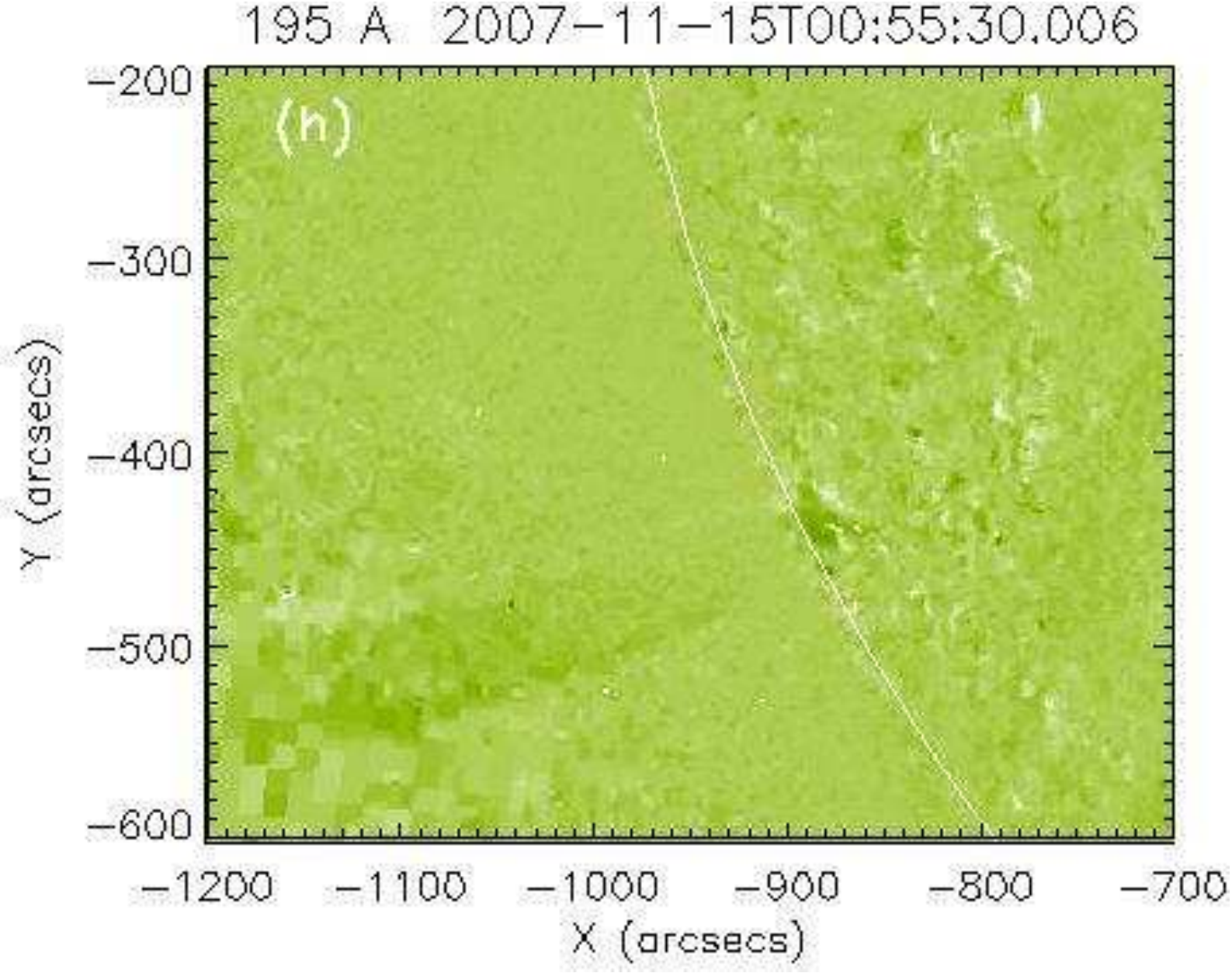} \\ 
\end{tabular}
\caption{Evolution of the jet seen by STEREO A at 304 \AA~ and 195\AA~on 15 November, 2007. For better visibility, there 195 \AA images are shown as difference images.}
\label{fig_5}

\end{figure}
\begin{figure}[t]
\vspace*{0mm}
\begin{tabular}{c}
\includegraphics[width=6.0 cm]{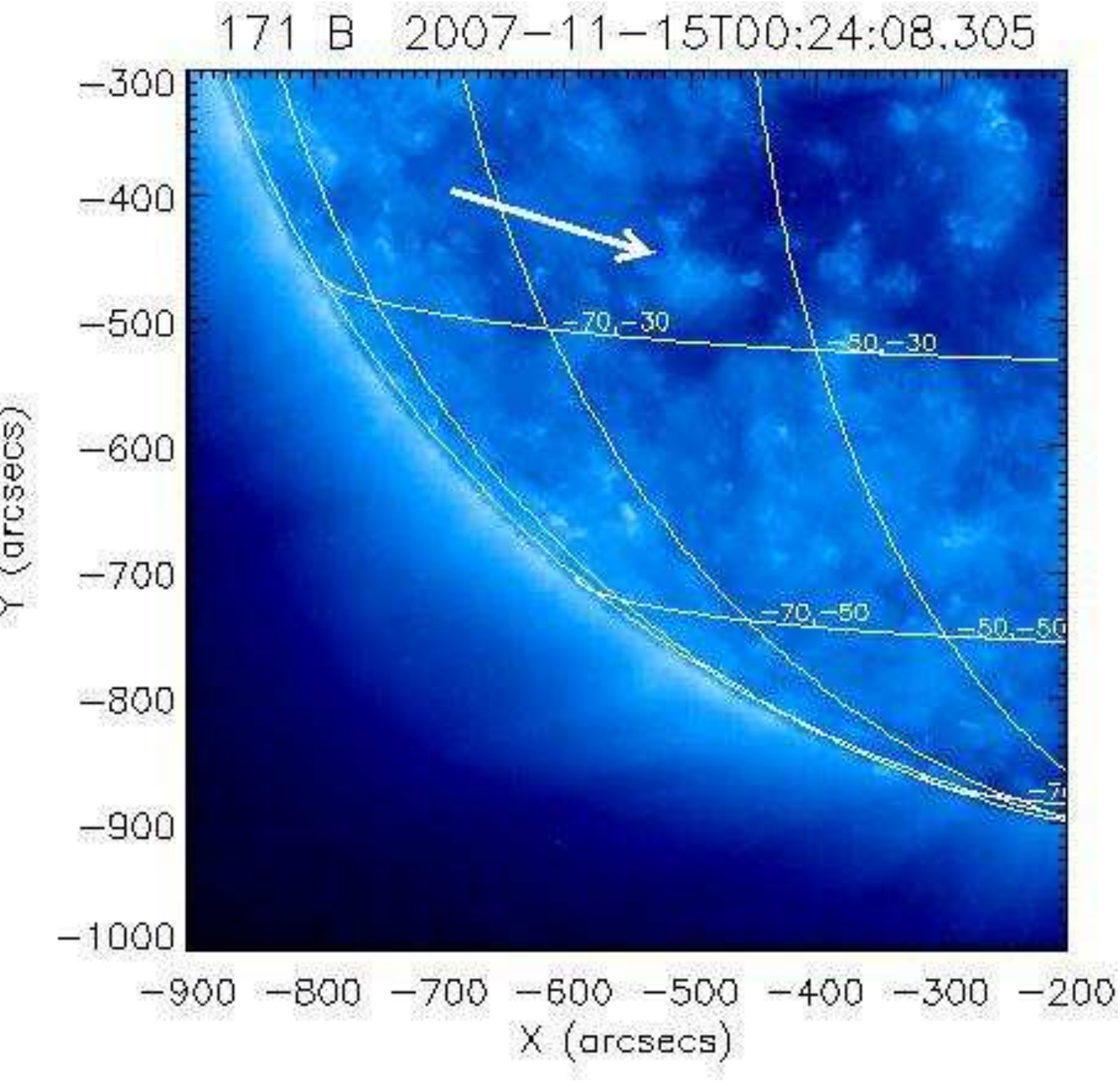}\\ 
\includegraphics[width=6.0 cm]{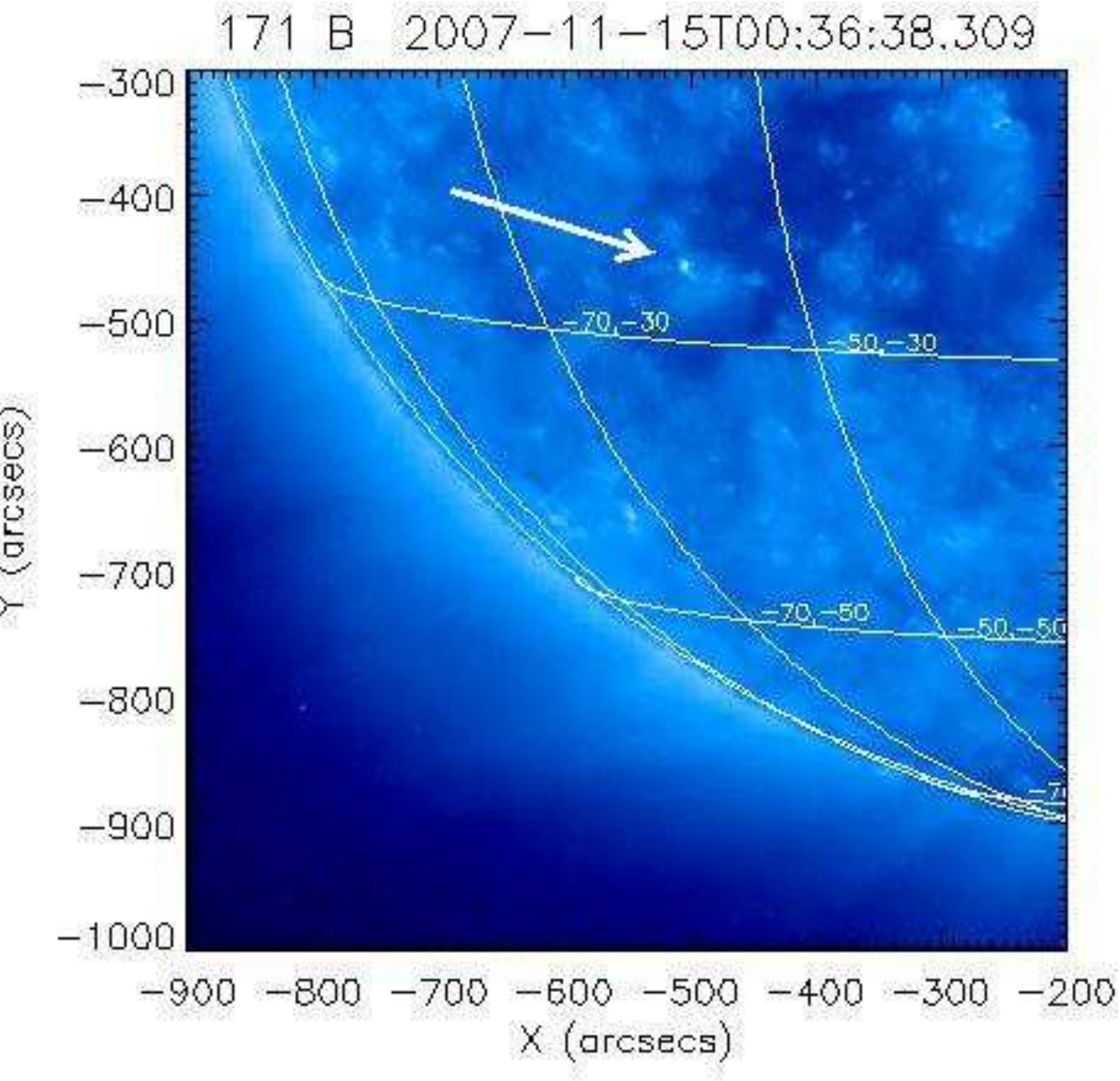}\\
\end{tabular}
\caption{Evolution of the bright point seen by STEREO B at 171 \AA~on 15 November, 2007, for the event shown in Fig. \ref{fig_5}}
\label{fig_6}
\end{figure}
The same coronal hole as in the case of the jet shown in Fig. \ref{fig_3} was the source region for another jet occuring four days later, on 15 November, 2007 (event N$^\circ$ 11). The peculiarity of this event is that it was visible by both STEREO spacecraft. The event occurred on the east limb in STEREO A; Fig. \ref{fig_5} shows its evolution seen by STEREO A at 304 \AA~and 195 \AA~as difference images. The ejection start time is about 00:26 UT, developing to the left side of the coronal hole at the East limb. First, a bright feature appeared on the lower solar atmosphere, followed by the eruption in which two arms can be seen, suggesting a helical magnetic field structure. A shift of the jet towards the southern direction is also seen, similar to the ``lambda'' event on 17 November, 2007, discussed by \citet{Nistico09}.
In that event the ejection seemed to have shifted from the initial position, where a bright point was found \citep{Nistico09, Moreno-Insertis08, Filippov09} to the other leg, and finally the jet became very narrow and collimated.
As can been seen from Fig. \ref{fig_5}f-g, this event also becomes narrow and collimated. A more dynamic impression of the jet evolution can be obtained from the on-line material.     
In Fig. \ref{fig_6} we show the same jet seen from the perspective of STEREO B at 171 \AA. Here the coronal hole appears clearly on the solar disk as a dark area within the brighter quiet Sun corona. The arrow, in the two images, marks the bright point, evolving at the coronal hole boundary. The bright point position is consistent with that of the jet seen by STEREO A. The visiblity of the bright point starts at 00:26  UT with its brightness increases in time. It is also visible in 195 \AA, while it is more difficult to see 304 \AA. The jet  is located in an area of the coronal hole which is less dark, which may be an indication of small-scale closed fields in this portion of the coronal hole. This observation suggests the possibility of magnetic reconnection between open magnetic field lines of the coronal hole and closed ones of quiet Sun-like regions ({\it e.g.} \citet{Madjarska04}). On the other hand, jets usually seem  to originate  as a consequence of reconnection between the open magnetic field of coronal holes and emerging magnetic bipoles. Further analysis , including the use of solar surface magnetograms, is needed to assess the magnetic topology in events like the one shown in Fig. \ref{fig_6}.

\begin{figure}[t]
\begin{center}
\vspace*{1 mm}
\begin{tabular}{l}
\includegraphics[width=8.5 cm]{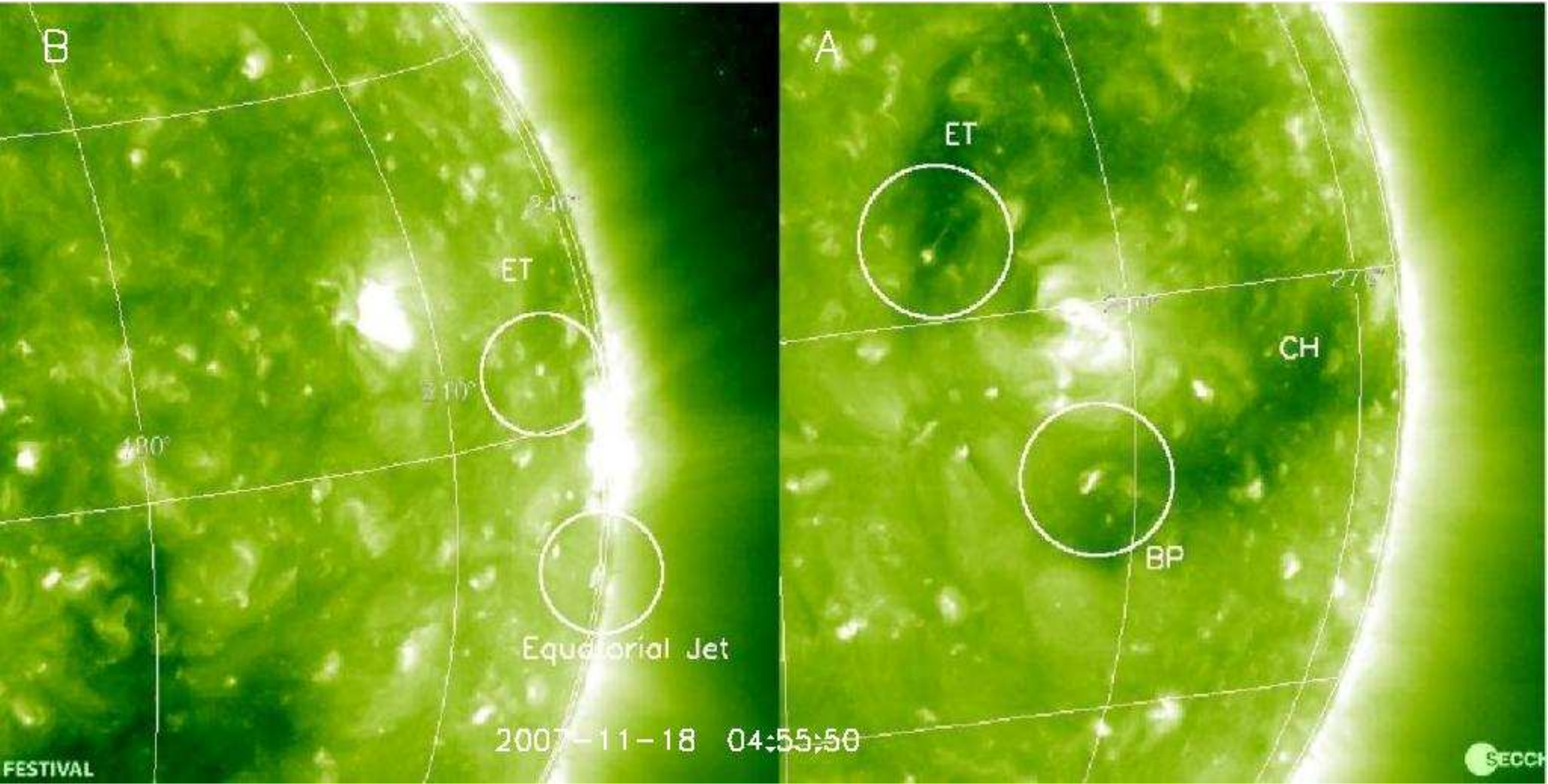} \\
\includegraphics[width=8.5 cm]{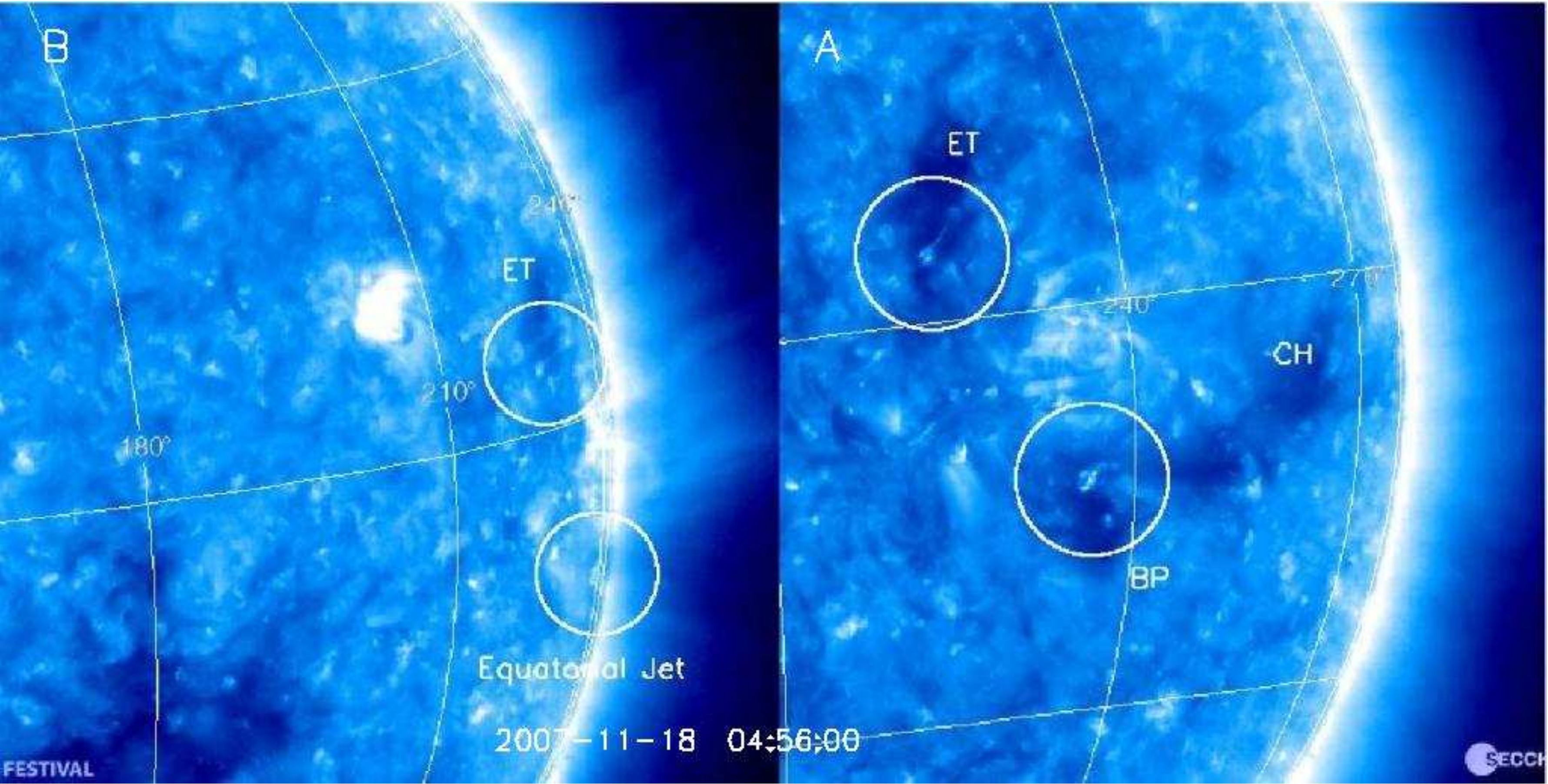} \\
\includegraphics[width=8.5 cm]{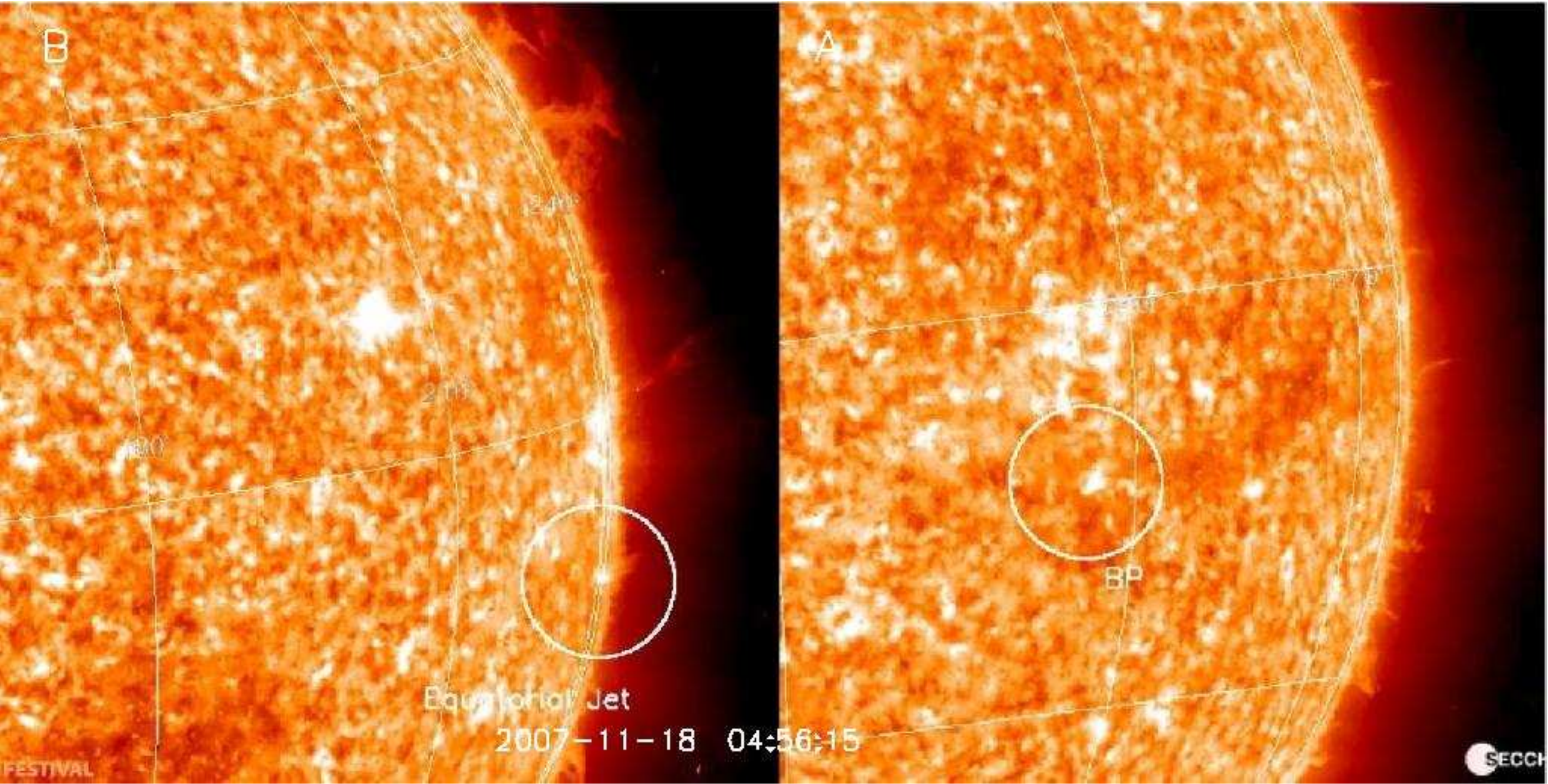} \\
\end{tabular}
\caption{STEREO B and A images at 195, 171 and 304 \AA~for the equatorial jet observed on 18 November, 2007.}
\label{fig_7}
\end{center}
\end{figure}
\begin{figure*}[t]
\begin{center}
\vspace*{1 mm}
\begin{tabular}{c c}
\includegraphics[width=6.5 cm]{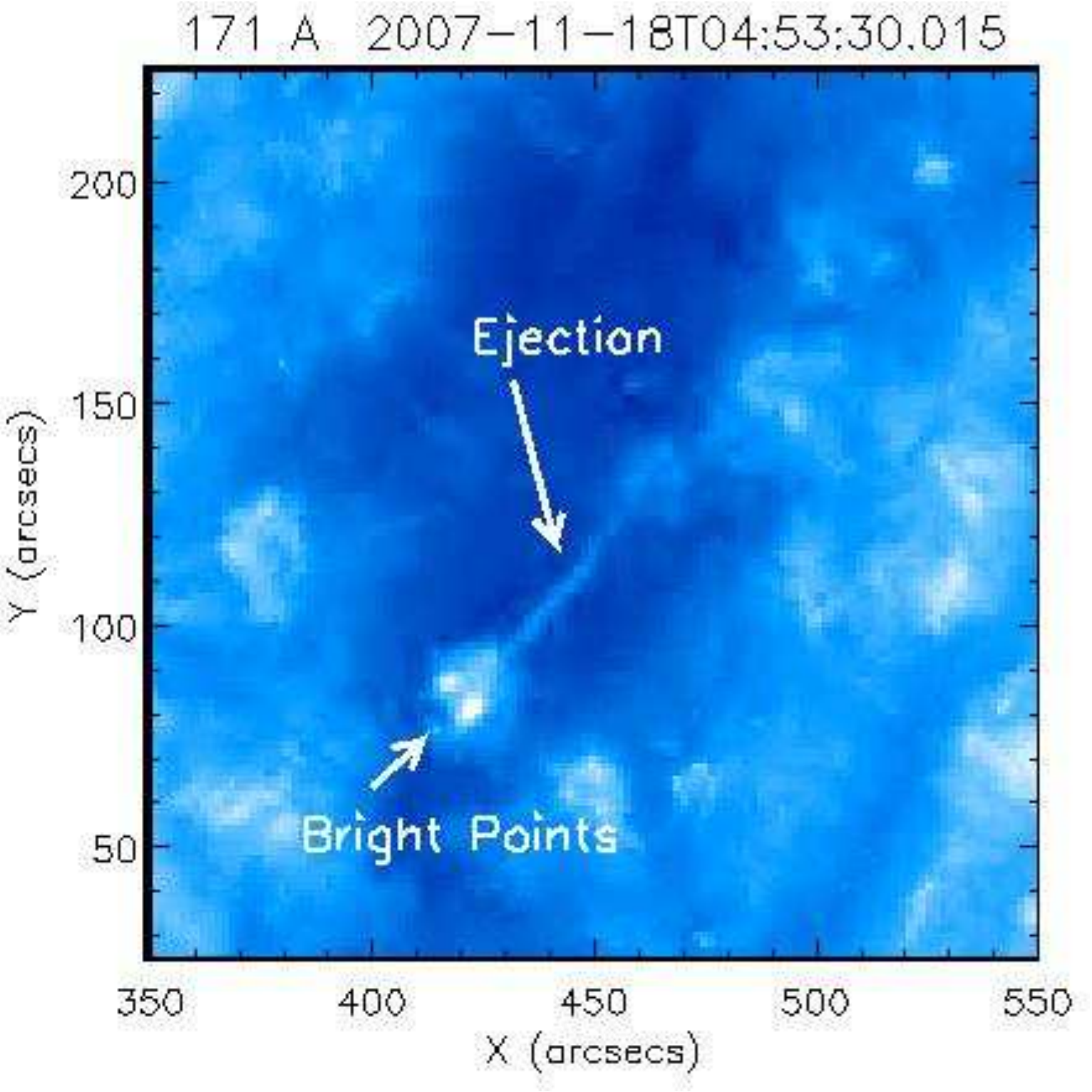} & 
\includegraphics[width=6.5 cm]{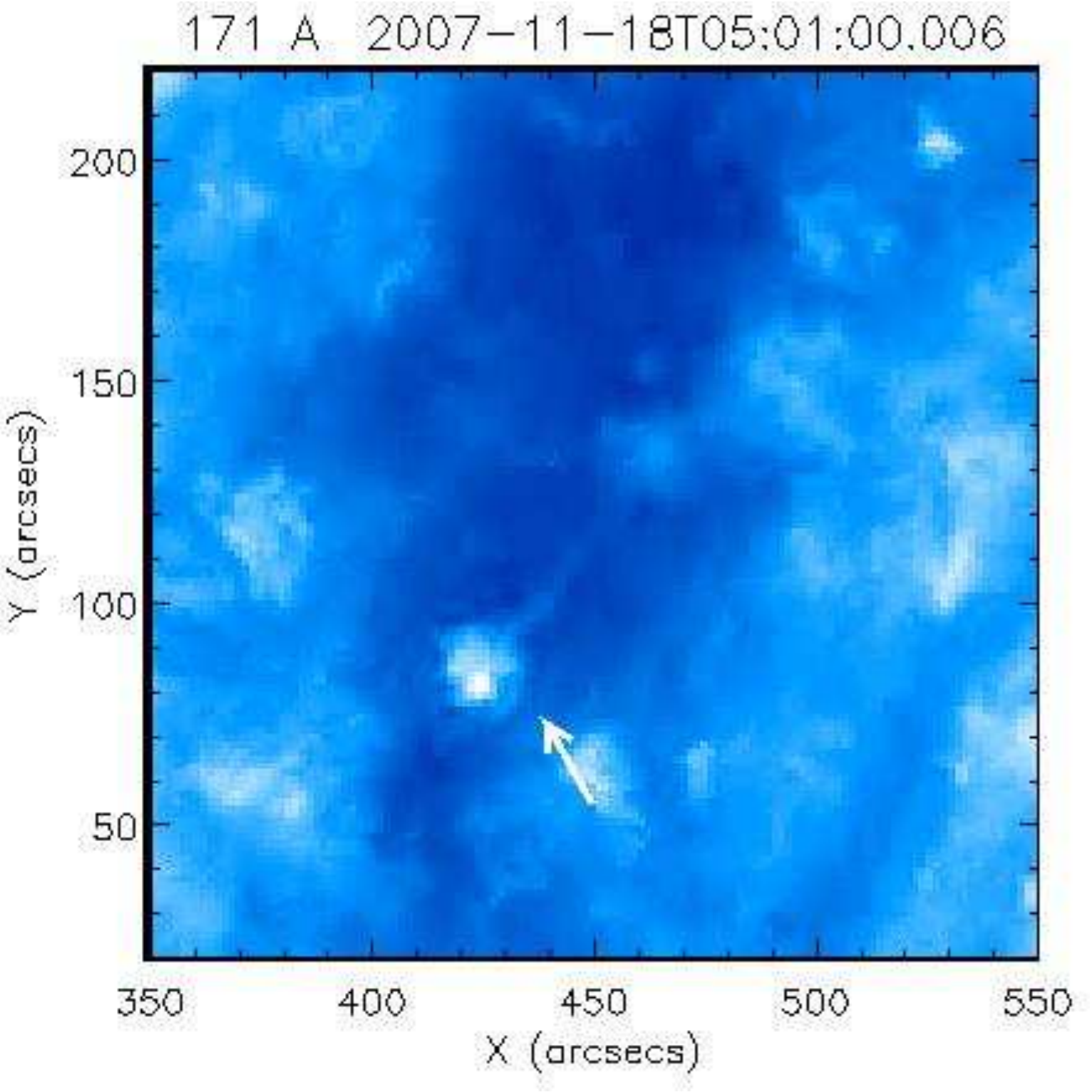} \\ 
\includegraphics[width=6.5 cm]{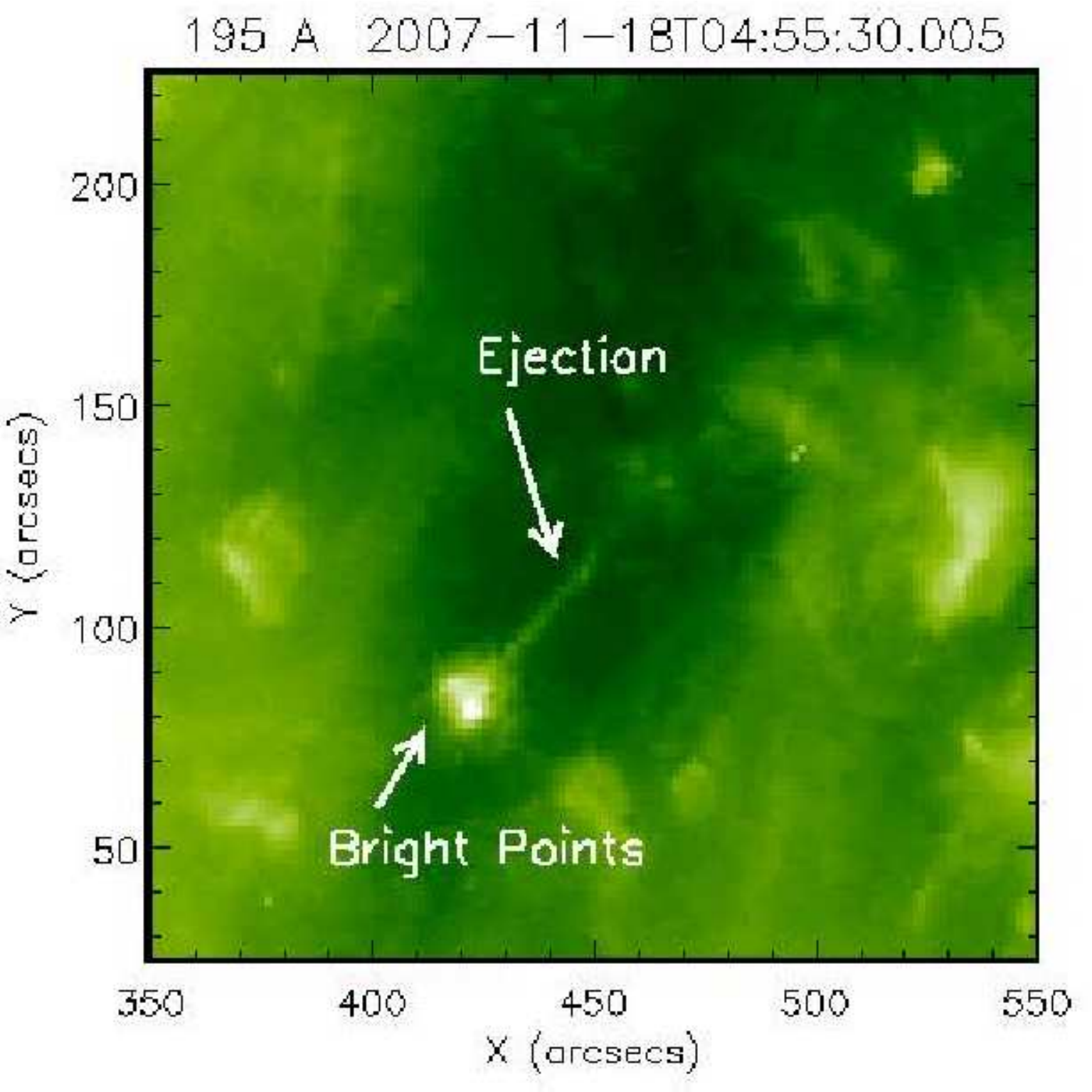} &
\includegraphics[width=6.5 cm]{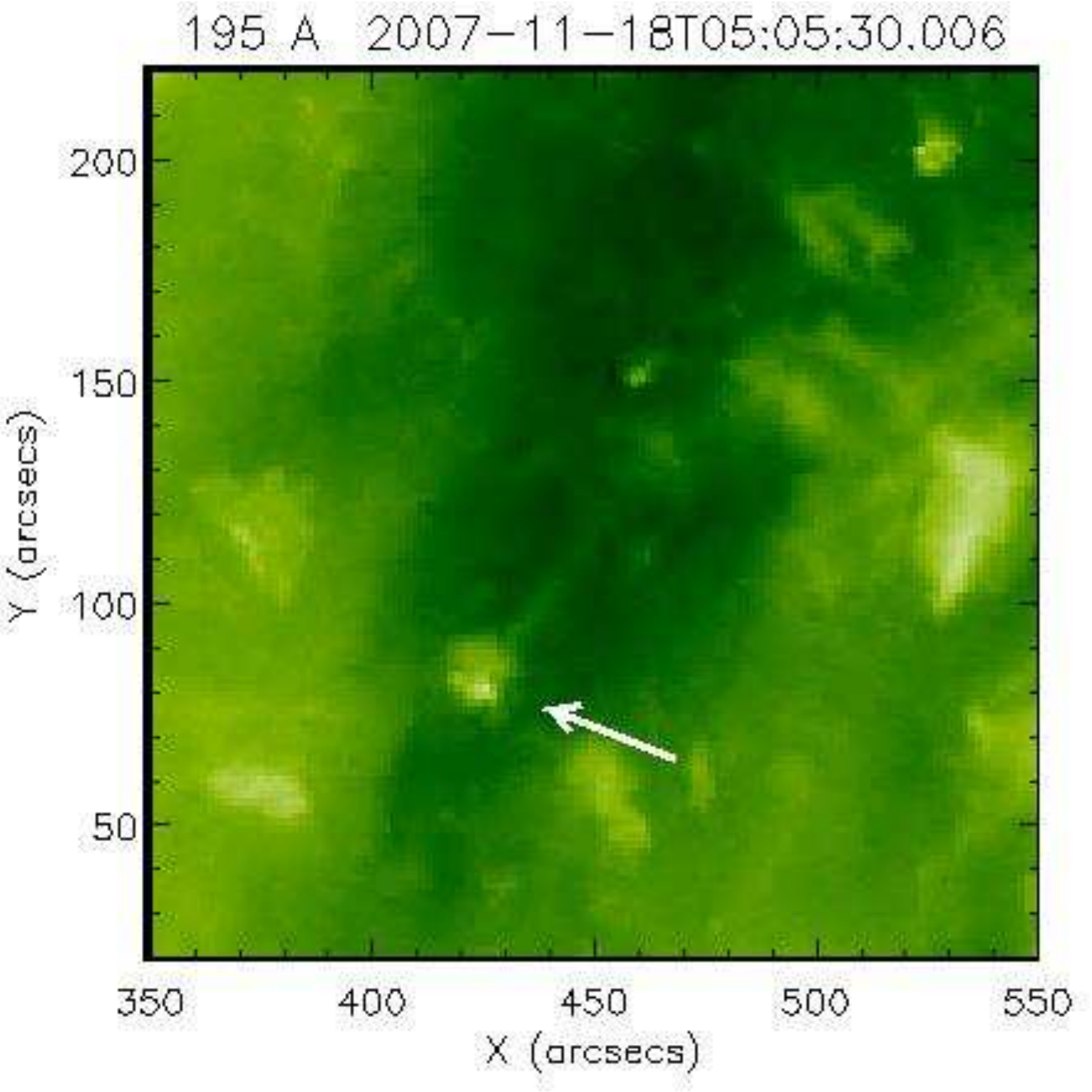}\\
\end{tabular}
\caption{A blow up of the ``Eiffel Tower'' jet shown in Fig. \ref{fig_7}.}
\label{fig_8}
\end{center}
\end{figure*}

Another example is provided by event N$^\circ$ 12 of Table \ref{table_1}, which happened on 18 November, 2007. In Fig. \ref{fig_7} we show images at 195, 171 and 304 \AA~for STEREO B (left) and A (right). At 195 and 171 \AA, one can clearly distinguish the presence of  coronal holes around an active region, visible at the limb with STEREO B. The coronal hole areas are darker than the quiet Sun areas but overall the corona appears very inhomogeneous, also reflected by the presence of many bright points. Through circles labeled by BP we have marked the presence of a bright point that increased its brightness in time as seen in STEREO A images, and for which the corresponding jet was visible at the limb with STEREO B. At 171 \AA~the emission pattern seems to initially involve two BPs, increasing in intensity, similar to an ``Eiffel-Tower'' (ET) shape.
Through the circle in the northern coronal hole, labeled by ET, we note a jet, evolving simultaneously to the one labeled BP. This jet exhibits an ET structure, characterized by 2 bright points at its footpoints, similar to the events discussed by \citet{Yamauchi04}. We note that this event is not listed in the Table \ref{table_1} because it did not show a counterpart in the COR 1 FOV. For the event labeled BP, at 304 \AA, the ejection was evident at the limb with STEREO B as a very bright footpoint, and visible with A inside the circle labeled  BP in Fig. \ref{fig_7}. Fig. \ref{fig_8} shows a blow up of the ET event at 171/195  \AA, which clearly reveals the ET structure.         

\begin{table*}[t]
\caption{List of equatorial coronal jet events identified from March to December 2007 from the STEREO EUVI and COR 1 observations.}
\vskip4mm
\centering
\scriptsize
\begin{tabular}{p{0.3 cm} p{1.5cm} p{0.5 cm } p{1.5 cm} p{2.5cm} p{1.5 cm} p{2.5 cm} p{2.5 cm}}
\tophline
\bfseries{N} & \bfseries{Date} &\bfseries{$\lambda(\AA)$}& \bfseries{COR1 A}  & \bfseries{EUVI A} & \bfseries{COR1 B}  & \bfseries{EUVI B} &  \bfseries{Position}  \\
\bfseries{$\Delta\phi$} &   &  &        &        &             &   \\ 
\middlehline
 1  & 2007-03-31 & 171 & 22:30 &        & 22:40  &             &   \\ 
    &            & 195 &       &        &        & 21:52-22:32 &  $\alpha(B)=120^\circ$ \\ 
 2.9&            & 284 &       &        &        &             &  $\beta(B)=118^\circ$ \\
    &            & 304 &       &        &        &             &   \\
\middlehline
 2  & 2007-04-07 & 171 & 16:40 & 15:59-16:41 & 16:40 & 15:59-16:41 &  \\
    &            & 195 &       & 15:22-16:42 &       & 15:42-16:32 & $\alpha(A)=70^\circ$ \\ 
 3.4&            & 284 &       & 16:00-16:40 &       & 16:00-16:40 & $\beta(A)=76^\circ$  \\
    &            & 304 &       &             &       &             &  \\
\middlehline
 3  & 2007-04-30 & 171 & 22:40 & 22:29-22:49& 22:40  &     ?     &   \\
    &            & 195 &       & 22:32-22:42&        &           &  $\alpha(A)=259^\circ$\\
 5.9&            & 284 &       &          &          &           &  $\beta(A)=241^\circ$   \\
    &            & 304 &       &          &          &           &   \\
\middlehline
 4  & 2007-05-01 & 171 & 21:30 &20:54-21:26 & 21:30  &20:54-21:26&    \\
    &            & 195 &       &            &        &           &  $\alpha(B)=255^\circ$  \\
 6.1&            & 284 &       &            &        &           &  $\beta(B)=249^\circ$  \\
    &            & 304 &       &21:01-21:21 &        &21:01-21:21&    \\
\middlehline 
 5  & 2007-05-16 & 171 & 13:05 &           & 13:25  &             &  \\
    &            & 195 &       &           &        &             & $\alpha(A)=104^\circ$ \\
 8.0&            & 284 &       &           &        &             & $\beta(A)=104^\circ$ \\
    &            & 304 &       &12:30-12:45&        &             &  \\
\middlehline
 6  & 2007-10-03 & 171 &   -   &   -       & 10:06  & 09:33-10:03 &  \\ 
    &            & 195 &       &           &        & 09:35-09:55 & $\alpha(B)=255^\circ$ \\
34.4&            & 284 &       &           &        &             & $\beta(B)=251^\circ$ \\
    &            & 304 &       &           &        & 09:36-10:06 &  \\
\middlehline
 7  & 2007-10-10 & 171 &  -    &    -     & 22:06  & 21:26-21:46  &  \\  
    &            & 195 &       &          &        & 21:35-21:45  & $\alpha(B)=85^\circ$\\
35.5&            & 284 &       &          &        &              & $\beta(B)=77^\circ$  \\
    &            & 304 &       &          &        & 21:26-21:56  &  \\
\middlehline
 8  & 2007-10-14 & 171 &   -   &    -     & 11:05  & 10:26-11:26  &  \\
    &            & 195 &       &          &        &              & $\alpha(B)=104^\circ$\\
36.1&            & 284 &       &          &        &              & $\beta(B)=94^\circ$  \\
    &            & 304 &       &          &        & 10:36-11:06  &  \\                  
\middlehline
 9  & 2007-10-15 & 171 &   -   &    -     & 17:46  & 17:13-17:43  & \\
    &            & 195 &       &          &        &  17:25-17:45 & $\alpha(B)=112^\circ$  \\
36.2&            & 284 &       &          &        &              & $\beta(B)=97^\circ$ \\
    &            & 304 &       &          &        &  17:16-17:46 &  \\
\middlehline
10  & 2007-11-11 & 171 &   -   &    -     &16:05   & 15:41-16:08  &  \\
    &            & 195 &       &          &        &  15:45-15:55 & $\alpha(B)=112^\circ$ \\
39.6&            & 284 &       &          &        &  15:46       & $\beta(B)=81^\circ$ \\
    &            & 304 &       &          &        &  15:36-16:16 &  \\
\middlehline
11  & 2007-11-15 & 171 & 01:05 &00:31-01:06&  -    &     -        &  \\
    &            & 195 &       &00:45-01:05&       &              & $\alpha(A)=116^\circ$ \\
40.1&            & 284 &       &          &        &              & $\beta(A)=115^\circ$ \\
    &            & 304 &       &00:36-01:06&       &              &  \\
\middlehline
12  & 2007-11-18 & 171 &  -    &  -       & 05:25  &              &  \\
    &            & 195 &       &          &        &              &  $\alpha(B)=256^\circ$\\
40.3&            & 284 &       &          &        &              &  $\beta(B)=265^\circ$  \\
    &            & 304 &       &          &        & 04:56-05:36 &   \\
\middlehline
13  & 2007-11-18 & 171 & 05:45 & 05:06-05:43 &  -  &   -         &    \\
    &            & 195 &       & 05:16-05:35 &     &             &  $\alpha(A)=124^\circ$\\
40.3&            & 284 &       &   05:26  &        &             &   $\beta(A)=126^\circ$  \\
    &            & 304 &       & 05:06-05:26 &     &             &   \\
\middlehline
14  & 2007-11-22 & 171 &  -   &    -      &11:25   & 10:56-11:33 &    \\
    &            & 195 &      &           &        & 11:05-11:35 &   $\alpha(B)=242^\circ$\\ 
40.8&            & 284 &      &           &        &             &   $\beta(B)=239^\circ$\\
    &            & 304 &      &           &        & 11:06-11:36 &   \\
\middlehline
15  & 2007-12-12 & 171 & 11:45&11:18-11:46& 11:45  &     -       &   \\
    &            & 195 &      &11:15-11:45&        &             &   $\alpha(A)=134^\circ$\\
42.9&            & 284 &      &           &        &             &   $\beta(A)=131^\circ$\\
    &            & 304 &      &11:16-11:46&        &             &   \\
\bottomhline
\end{tabular}
\label{table_1}
\end{table*}

\conclusions
In this paper we present observations of equatorial coronal hole jets identified in images of the STEREO EUVI and COR 1 telescopes, of the SECCHI instruments suites on board the twin STEREO spacecraft. The observation time and location of these jets seen with STEREO A and B is provided through Table \ref{table_1}. From a detailed inspection of the observations derived from the two perspectives, we can show that the bright points previously observed in association with polar jets, corresponded to the footpoint of the jets, and that the collimated plasma beam became visible only when it appeared as bright structure at the solar limb against a dark background. For event N$^\circ$ 10 of Table \ref{table_1} we determined its kinematics in detail. The jet accelerated impulsively to a speed of almost 200 km/s. Further, the plasma of the emitting region exhibited a downward acceleration of about 0.11 km s$^{-2}$, which is less than the solar gravitational force. This suggests that other forces, like pressure gradients or electromagnetic forces, are acting on the jet material. For the events (N$^\circ$ 11, 12) of Table 1 we distinguished ``lambda'' and ``Eiffel Tower'' topologies, similar to those observed for polar jets. However, for most events, a straightforward classification of their morphology is not possible because of the difficult visibility of the jets observed at low latitudes mostly due to the presence of ambient brighter coronal structures, such as helmet streamers. The lifetime of the equatorial jets was found to be of the order of 20-40 minutes, {\it i.e.}comparable to those of polar jets. The inferred velocities were of the order of 200 km s$^{-1}$ for those events for which length and thickness could be reasonably well measured, being also consistent with the values derived for polar coronal jets. From these results, but taking caution into account because only very few events could be investigated, we conclude that there is no differences between polar and equatorial coronal hole jets in their origin and basic characteristics.

\begin{acknowledgements}
It is a pleasure to thank all the STEREO staff, without which this work would not have been possible. G. N. acknowledges support from an Erasmus grant during his stay in Goettingen and support from the Regione Calabria of Italy for partecipating to the STEREO-3/SOHO-22 workshop. V. B acknowledges the support of the project Stereo/Corona by the German Bundesministerium f\"{u}r Bildung und Forschung through the deutsche Zentrum f\"{u}r Luft-und Raumfahrt e.V. (DLR, German Space Agency) as a collaborative effort with the Max-Planck-Institut f\"{u}r Sonnensystemforschung (MPS) under grant 50 0C 0904. Stereo/Corona is a science and hardware contribution to the optical image package SECCHI, developed for the NASA STEREO mission launched in 2006. The SECCHI data used here were produced by an international consortium of the Naval Research Laboratory (USA), Lockheed Martin Solar and Astrophysics Lab (USA), NASA Goddard Space Flight Center (USA), Rutherford Appleton Laboratory (UK), University of Birmingham (UK), Max-Planck-Institut for Solar System Research (Germany), Centre Spatiale de Li\'ege (Belgium), Institut d’Optique Th\'eorique et Appliqu\'ee (France), and Institut d'Astrophysique Spatiale (France). The research leading to these results has received funding from the European Community's Seventh Framework Programme (FP7/2007-2013) under the grant agreement n$^\circ$ 218816 (SOTERIA project, \url{www.soteria.eu}). G. Z. was supported in part by the Italian INAF and by the Italian Space Agency, contract ASI n. I/015/07/0 ``Esplorazione del Sistema Solare''.
Some images are produced by FESTIVAL, collaborative project managed by IAS and supported by CNES, which is  a SolarSoftware package for processing and manipulation of solar image data.   
\end{acknowledgements}

\end{document}